\documentclass{mn2e}
\usepackage{epsfig,natbib2,natbibmnfix}
\usepackage{amsmath}
\usepackage{mathrsfs}
\usepackage{subfigure}
\usepackage{url}
\usepackage[dvips]{color}
\usepackage{multicol}

\newcommand{\be}{\begin{equation}}
\newcommand{\ee}{\end{equation}}

\def\ltsima{$\; \buildrel < \over \sim \;$}
\def\simlt{\lower.5ex\hbox{\ltsima}}
\def\gtsima{$\; \buildrel > \over \sim \;$}
\def\simgt{\lower.5ex\hbox{\gtsima}}

\def\del#1{{}}

\def\msun{{\,{\rm M}_\odot}}
\def\tsun{{\,{\rm t}_\odot}}

\title[Superbubble-driven accretion flows]
{Growing galaxies via superbubble-driven accretion flows}

\author[Alexander Hobbs, Justin Read, Andrina Nicola]
       {\parbox{18cm}{Alexander Hobbs$^{1}$, Justin Read$^{2}$, Andrina Nicola$^{1}$}\vspace{0.3cm}\\
\noindent $^{1}$Institute for Astronomy, ETH Zurich $^{2}$University of Surrey, UK}

\begin{document}

\maketitle

\begin{abstract}
We use a suite of cooling halo simulations to study a new mechanism for rapid accretion of hot halo gas onto star-forming galaxies. Correlated supernovae events create converging `superbubbles' in the halo gas. Where these collide, the density increases, driving cooling filaments of low metallicity gas that feed the disc. At our current numerical resolution ($\sim 20$ pc) we are only able to resolve the most dramatic events; these could be responsible for the build-up of galaxy discs after the most massive gas-rich mergers have completed (redshift $z\simlt 1$). As we increase the numerical resolution, we find that the filaments persist for longer, driving continued late-time star formation. This suggests that SNe-driven accretion could act as an efficient mechanism for extracting cold gas from the hot halo, driving late time star formation in disc galaxies. We show that such filament feeding leads to a peak star formation rate (SFR) of $\sim 3$\,M$_\odot$\,yr$^{-1}$, consistent with estimates for the Milky Way. By contrast, direct cooling from the hot halo (`hot-mode' accretion, not present in the simulations that show filament feeding) falls short of the SNe-driven SFR by a factor of $3-4$, and is sustained over a shorter time period. The filaments we resolve extend to $\sim 50$\,kpc, reaching column densities of $N \sim 10^{18}$ cm$^{-2}$. We show that such structures can plausibly explain the broad dispersion in Mg II absorption seen along sight lines to quasars. Our results suggest a dual role for stellar feedback in galaxy formation, suppressing hot-mode accretion while promoting cold-mode accretion along filaments. This ultimately leads to more star formation, suggesting that the positive feedback effect outweighs the negative -- at least in the models presented here. Finally, since the filamentary gas has higher angular momentum than that coming from hot-mode accretion, we show that this leads to the formation of substantially larger gas discs. We are only just beginning to be able to resolve such processes in our simulations.
\end{abstract}

\renewcommand{\thefootnote}{\fnsymbol{footnote}}
\footnotetext[1]{E-mail: {\tt ahobbs@phys.ethz.ch}}

\section{Introduction}
The means by which gas gets into galaxies is currently an open question. In particular, the supply of cold ($< 10^4$ K) star-forming gas in late-type galaxies at $z \simlt 1-2$ is a puzzle. At higher redshift, cosmological simulations predict a dominant mode of accretion where gas flows along overdense dark matter (DM) channels without being shock heated to the virial temperature \citep[e.g.,][]{DekelEtal09}. This allows cold gas to penetrate deep within the potential well of the DM halo. However, this `cold mode' phase transitions to a `hot mode' by $z \sim 1-2$ that forms and maintains a hot, rarefied corona around the galaxy \citep{KeresEtal2009}.

The hot mode is difficult to reconcile with star formation histories (SFHs) of late-type galaxies. Typically, there is a great deal of star formation that continues at a near-constant/slightly declining rate for $\sim 9$ Gyrs after the cold mode has ended \citep{Rocha-PintoEtal2000, AumerBinney2009}. Since the gas consumption timescale -- the rate at which gas is converted into stars -- is very short compared with the overall SF trend across the lifetime of the galaxy, the gas accretion rate must balance the SFR. This means that cold star-forming gas must continually be able to make its way into the galactic disc throughout this time. In theory, the more massive cosmological coronae can provide more than enough gas to fuel the level of star formation seen in Milky Way-type galaxies. However, the gas must first cool in order to accrete onto the disc.

Until recently, it was thought that this could be achieved through sudden and distributed cooling from the hot corona in the form of cold clouds, brought about by thermal instability \citep{MallerBullock2004, KaufmannEtal2006, KaufmannEtal2007, KaufmannEtal2009}. However, this picture has since been shown to be problematic \citep{BinneyEtal2009, JoungEtal2012, HobbsEtal2013}. Firstly, the conditions for both linear and non-linear thermal instability are not met in typical stratified coronae. Rather, such behaviour, when seen in smoothed particle hydrodynamics (SPH) simulations, owes to an inability to correctly treat the mixing of different fluid phases \citep{HobbsEtal2013}. Secondly, if coronae were thermally unstable, one would expect to find multiple cold clouds distributed in a uniform fashion throughout. However, ultra-sensitive HI surveys have been unable to find HI clouds at significant distances from galaxies, or in galaxy groups \citep{PisanoEtal2007, IrwinEtal2009}. Wherever massive HI complexes \emph{are} seen, they are typically associated with stars in the interstellar medium (ISM) of a galaxy \citep{DoyleEtal2005, SaintongeEtal2008}.

Simple physical arguments point to the hot corona being unable to cool efficiently on its own. For a late-type galaxy, a corona created from cosmological sources of accretion would have a long cooling time, perhaps $\sim$ a few-$10$ Gyrs \citep{HodgesBregman2013}. Moreover, early-type galaxies, which have readily detectable X-ray emitting hot coronae \citep{FormanEtal1985, MorgantiEtal2006}, lack any sign of significant cold gas reservoirs or ongoing star formation.

Thus, to explain the apparent cooling of the corona around star-forming disc galaxies, a mechanism from an alternative source must be invoked. An external actor in the form of cosmological filaments from large scales is ruled out based on the $z \simlt 1-2$ accretion mode transition as well as the fact that such a mechanism would also have to be effective in early-type galaxies. In simulations, when cold cosmological filaments \emph{are} seen in the vicinity of massive halos ($M > 3 \times 10^{11} \msun$) or below the redshift transition, they typically get disrupted well before they can reach the galactic disc \citep{KeresEtal2005, FernandezEtal2012}.

We are therefore left to posit an \emph{internal} actor for our cooling mechanism. The star-forming disc that provides the evidence for sustained cold gas inflow may also be the cause of it. In a series of papers, \cite{MarinacciEtal2011, MarascoEtal2012, MarascoEtal2013} explore the idea that cold gas can be extracted from the hot halo by a galactic fountain effect driven by supernovae (SNe) in the disc. The SNe send disc gas into the lower regions of the hot halo, where it mixes with hot gas, entraining and cooling it. The parcel of gas then falls back down onto the disc, providing it with new fuel for further star formation. Such a process is a form of \emph{positive feedback} from SNe, and allows for an accretion rate similar to the Galactic SFR and reasonable consistency with the kinematics of extraplanar HI gas in the Milky Way \citep{MarascoEtal2012, 2008A&A...487..951K}.

In this paper we propose an alternative positive feedback process, arising out of a star-forming proto-galaxy at the centre of a massive hot halo. Our mechanism is distinct from the model of \cite{MarinacciEtal2011} in that it links halo gas at large scales ($\sim 50$ kpc) to fuel for SF in the galactic disk through an accretion channel. The net accretion rate onto the disk for the duration of the process is substantial ($0.1-3 \msun$ yr$^{-1}$), and gas is drawn preferentially from a low-metallicity phase. The accretion channel is created by colliding flows caused by `hot bubbles' driven by SNe, promoting cooling through overdensity and thermal instability. In the \cite{MarinacciEtal2011} model, on the other hand, a `seed' of cold, high-metallicity gas from the disc is required to mix with the halo gas and lower the cooling time of the mixture. This is an important distinction, since the latter mechanism is dependant on the chemical properties of the `seed' gas, while ours has only a density requirement. However, we believe that both mechanisms could well act in tandem in real galaxies.

The behaviour of individual clouds, presumed to be formed from inflowing filamentary gas, has been investigated by \cite{JoungEtal2012a}. Typically these clouds stall in cooling at just under $10^4$ K, which coincides with a sudden reduction in cooling efficiency in standard cooling curves (a sharp drop to the left of the peak of the cooling curve). Due to their nature as single clouds rather than cloud or filamentary complexes, they are in general disrupted and mix back in with the ambient gas before reaching the galactic disk. As a result, their capacity for providing the disk with a supply of cold gas is diminished. As we will see in the current paper, however, our model of extended filaments provides a channel for $\sim 10^4$ K gas to flow through without being disrupted or hindered by the surrounding hotter gas.

Such compression regions between SNe-generated outflows have been studied numerically in the context of star-forming regions on smaller scales \citep{NtormousiEtal2011}, with the authors finding a strongly-enhanced tendency for the gas in the collision region to form structures such as clouds and filaments. These structures dominate the cold gas budget for the simulation and show substantial internal motions. Our accretion flow mechanism is therefore related but occurs at larger scales within a broader galaxy formation context, with the gas flowing primarily down a steep potential gradient to the central galaxy. Observationally, there is a growing body of evidence for such filamentary structures both within star-forming molecular cloud complexes \citep{LeeChen2009, LuongEtal2011, PonEtal2014} and on galactic scales \citep{CecilEtal2002}. More generally, outflows powerful enough to create superbubbles of the size (or larger) seen in the current work have been observed in Lyman-$\alpha$ absorption \citep[e.g.][]{ErbEtal2012}.

This paper is organised as follows. In Sections 2 \& 3, we describe our numerical method and initial conditions. In Sections 4-7, we describe our sub-grid physics implementation. In Section 8, we present our results. In Section 9-1, we discuss these results and their implications for galaxy formation; and we compare and contrast our findings with previous works in the literature. Finally, in Section 13 we present our conclusions.

\section{Computational method}\label{sec:numerics}

For the simulations, we use the smoothed particle hydrodynamics (SPH) code GADGET-3 \citep{Springel05}, with the SPHS algorithm \citep{2009arXiv0906.0774R, 2012MNRAS.422.3037R} to capture the mixing of different fluid phases. We employ a cubic spline (CS) kernel with 96 neighbours\footnote{It should be noted that 96 is the minimum neighbour number for the SPHS 2nd derivatives to be robust \citep{2012MNRAS.422.3037R}. In \cite{HobbsEtal2013} we demonstrated that CS96 gives excellent agreement with the more accurate HOCT442 for this problem, while at the same time providing more spatial resolution for the same numerical cost.}. Smoothing lengths for gas particles are spatially and temporally variable, adjusted to give a fixed gas mass inside the kernel.

The setup in terms of initial conditions, radiative cooling, and star formation and feedback, is the same as described in \cite{HobbsEtal2013}. We summarise it briefly in the next few subsections, but for a more detailed overview the reader is directed to the original paper \citep{HobbsEtal2013}.

\subsection{Initial conditions}\label{sec:ic}

For our initial condition, we use a live dark matter (DM) spherical halo of collisionless particles together with a cooling spherical halo of gas particles, with each component initially relaxed for many dynamical times with an adiabatic equation of state (EQS) to remove Poisson noise. Both the DM and gas distributions follow a Dehnen-McLaughlin model \citep{DehnenMcLaughlin2005} of the form:
\begin{equation}
\rho(r) \propto \frac{1}{(r/r_s)^{7/9} \left[1 + (r/r_s)^{4/9} \right]^6}
\label{eq:rho}
\end{equation}
with $1.5 \times 10^{11} \msun$ in the gas and $1.5 \times 10^{12} \msun$ in the DM. The scale radius for both halos is $r_s = 40$ kpc, and they are truncated at a virial radius of $r_t = 200$ kpc.

At the start of the simulation after relaxation the gas is in hydrostatic equilibrium, with a temperature profile given by the relation:
\begin{equation}
T(r) = \frac{\mu m_p}{k_{\rm B}}\frac{1}{\rho_{\rm gas} (r)} \int_r^\infty \rho_{\rm gas} (r) \frac{GM(r)}{r^2} \, dr
\end{equation}
where $\mu$ is the mean molecular weight, $k_{\rm B}$ is the Boltzmann constant, $\rho_{\rm gas} (r)$ is the radial gas density profile and $M(r)$ is the enclosed mass of both components within a radius $r$. The gas is assigned a rotational velocity field about the $z$-axis whereby the specific angular momentum profile follows a power-law \citep{BullockEtal2001b, KaufmannEtal2007} such that:
\begin{equation}\label{eq:ic}
j_{\rm gas} \propto r^{1.0}
\end{equation}
and normalised by a rotation parameter $\lambda = 0.038$, defined by:
\begin{equation}
\lambda = \frac{j_{\rm gas} \vert E_{\rm dm} \vert^{1/2}}{GM_{\rm dm}^{3/2}}
\end{equation}
where $E_{\rm dm}$ and $M_{\rm dm}$ are the total energy and mass of the DM halo. This normalisation implicitly assumes negligible angular momentum transport between the DM halo and the gas \citep{KaufmannEtal2007}. For the DM halos, we employed variable particle masses in order to minimise computational expense. The implementation of this is described in Section 3.1 in \cite{ColeEtal2011}. The binned density profile for the ICs can be seen in Figure \ref{fig:profile}, black line.

In the context of galaxy formation, the setup of the gaseous halo is designed to approximate (in a spherically-symmetric fashion) the main stage of (smooth) disc build-up that might arise after the last major merger \citep{AbadiEtal2003, Sommer-LarsenEtal2003, GovernatoEtal2004}. The initial behaviour of our simulation, due to the inclusion of supernovae feedback in the sub-grid model, is akin to a strong central starburst.

We note that the choice of non-cosmological ICs was deliberate in order to more easily isolate the effect of the SNe-driven accretion from cosmological accretion and/or mergers. We will study SNe accretion in a cosmological context in a forthcoming paper.

\subsection{Radiative cooling}\label{sec:cooling}

The gas is allowed to cool radiatively down to a cooling floor of $T_{\rm floor} = 100$\,K (although we note that almost all of the gas in the simulation remains at least an order of magnitude above this for the entire duration). We use a cooling curve that follows \cite{KWH1996} above $10^4$\,K, assuming primordial abundance, and \cite{MashchenkoEtal2008} below $10^4$\,K assuming solar abundance. This crudely models a low metallicity cooling halo that rapidly reaches $\sim$ solar abundance in cooling star forming regions. In addition to the cooling floor, we ensure that gas cannot cool beyond the point at which the Jeans mass for gravitational collapse becomes unresolved within our simulation. The mass resolution of our SPHS simulations is given by $M_{\rm res} = N_{\rm res} m_{\rm gas}$, where $m_{\rm gas}$ is the mass of a gas particle and $N_{\rm res}$ is the number of gas particles that constitutes a resovable mass, i.e., a single resolution element \citep[see, e.g.][]{BateBurkert1997}. We choose a value of $N_{\rm res} = 128$ \citep[see][]{ReadEtal2010, HobbsEtal2013, 2012MNRAS.422.3037R, DehnenAly2012}. This `dynamic' cooling floor effectively ensures that the Jeans mass is always resolved within our simulation. For a given mass element $M_{\rm res}$ we can write a Jeans density, namely,
\begin{equation}
\rho_J = \left(\frac{\pi k_b T}{\mu m_p G}\right)^3 \left(\frac{1}{M_{\rm res}}\right)^2
\label{eq:polytrope}
\end{equation}
where $k_b$ is the Boltzmann constant; $G$ is Newton's constant; and $\mu$ is the mean molecular weight.

We use equation \ref{eq:polytrope} to define a polytropic equation of state $P = A(s) \rho^{4/3}$ such that gas is not allowed to collapse (for a given temperature) to densities higher than given by $\rho_J$. We identify gas that lies on the polytrope as `star-forming', allowing it to form stars above a fixed density threshold (see Section \ref{sec:sf}). The inclusion of the polytrope prevents artificial fragmentation below our resolvable Jeans mass.

\subsection{Star formation and feedback}\label{sec:sf}

Star formation (SF) is modelled in a sub-grid fashion according to observational constraints on the SF rate and efficiency. We allow gas that lies on the polytrope to form stars above a fixed density threshold of $100$ atoms cm$^{-3}$, in locally converging flow with $\nabla \cdot \vec{v} < 0$, with an efficiency of 0.1 as per observations \citep[e.g.,][]{LadaLada2003} of giant molecular clouds (GMCs). The star formation rate follows the Schmidt volume density law for star formation \citep{Schmidt1959}, namely:
\begin{equation}
\rho_{\rm SFR} \propto \rho_{\rm gas}^{1.5}
\end{equation}
which is implemented in the simulation by employing the dynamical time as the relevant SF timescale, i.e.:
\begin{equation}
\frac{\text{d} \rho_*}{\text{d}t} = \epsilon \frac{\rho_{\rm gas}}{t_{\rm dyn}}
\end{equation}
where $\epsilon = 0.1$ is the SF efficiency and $t_{\rm dyn} = (4 \pi G \rho)^{-1}$, with $\rho$ being the density of the `star-forming' gas particle.

In all the simulations, we include feedback from supernovae (SNe), which takes the form of an injection of thermal energy from the star particle into nearby gas particles. Each star particle is representative of a stellar distribution, and so we integrate over a Salpeter IMF \citep{Salpeter1955} between $8 \msun$ and $100 \msun$ in order to determine the number of SNe that explode at any particular time. We treat only type II SNe, with the energy of each supernova event set to $E_{\rm SN} = 10^{51}$ ergs. To determine the time of injection we use the standard relation:
\begin{equation}
\frac{t_{\rm MS}}{\tsun} \sim \left(\frac{M}{\msun}\right)^{-2.5}
\end{equation}
to obtain an approximate main sequence time $t_{\rm MS}$ after the initial formation, at which all of calculated energy from the exploding SNe is injected at once. We couple the feedback energy to a given \emph{mass} of gas rather than just the neighbours of the star particle - this mass is set to be our resolvable mass $M_{\rm res}$ (see Section \ref{sec:cooling}). In doing so we ensure that both the star formation and the feedback is resolved. The actual injection of energy is convolved with a CS kernel to ensure a smoothing of the thermal coupling over the mass scale receiving the supernova feedback.

The polytropic pressure floor, and the fixed SF density threshold together define our sub-grid modelling strategy for star formation; gas that is allowed to form stars is (i) Jeans unstable (ii) at typical star-forming GMC densities (iii) numerically resolved (iv) in a converging $\nabla \cdot \vec{v} < 0$ region and (v) consistent with the other parameters of our setup e.g. the radiative cooling range.

\subsection{Numerical convergence}\label{sec:numconv}

Our approach, above, constitutes just one way in which we could build a `sub-grid' model for star formation and feedback. Our strategy here is to keep the star formation density threshold $n_{\rm th}$ fixed and then attempt to converge on this density by raising the numerical resolution. An alternate strategy is to scale $n_{\rm th}$ with the resolution. From equation \ref{eq:polytrope}, given a temperature floor and mass resolution we could define a maximum density which would be a natural choice for $n_{\rm th}$: 

\begin{equation} 
\left(\frac{n_{\rm th}}{{\rm atoms}\,/{\rm cm}^3}\right) = 15.2 \left(\frac{T}{1000\,{\rm K}}\right)^3 \left(\frac{10^4\,{\rm M}_\odot}{m_{\rm part}}\right)^2
\label{eqn:denthreshsimple}
\end{equation}
which suggests that even our highest resolution simulation in this paper ($m_{\rm part} = 4\times 10^4$\,M$_\odot$) is some way away from resolving $n_{\rm th} = 100$ \; atoms/cm$^3$. However, equation \ref{eqn:denthreshsimple} assumes that gas is supported only by thermal rather than turbulent pressure and/or energy input from feedback. Furthermore, it remains to be seen whether convergence will be faster when scaling $n_{\rm th}$ with resolution, or holding it fixed at some large value.

Similarly, we may worry about how the SNe energy is deposited in the gas surrounding a star forming region. Here, we thermally dump the energy, spreading it over neighbouring particles weighted by the kernel smoothing function. If we assume that the energy will be deposited on a scale set by the smallest smoothing length and at a density $\sim n_{\rm th}$, then we can simply calculate the temperature the gas will rise to as:

\begin{equation} 
T_c \sim \frac{2}{3 k_b}\frac{E_{\rm SNe}}{n_{\rm th} V}
\label{eqn:TcSNe}
\end{equation}
where $V$ is the volume over which the energy is dumped; $E_{\rm SNe} = N_{\rm SNe} 10^{51}$\,erg is the total supernovae energy; and $N_{\rm SNe}$ is the number of SNe that go off over a `main sequence' time $t_{\rm MS}$. 

Putting in numbers from the simulations we present in \S\ref{sec:accretionflow}, this gives: $T_c \sim [2 \times 10^4, 3 \times 10^5, 6 \times 10^6]$\,K for the three different mass resolutions we consider here, respectively. This is interesting because it means that in our lowest resolution run, the gas temperature after SNe injection will lie close to the peak of the cooling curve (see \S\ref{sec:cooling}), causing the SNe energy to rapidly cool away. By contrast, our highest resolution simulation pushes gas to a temperature well beyond the peak of the cooling curve, allowing significantly more work to be done before the gas can cool. Such considerations have led many authors to either switch off cooling in the ejecta gas \citep[e.g.][]{2006MNRAS.373.1074S}, or to inject the SNe as kinetic rather than thermal energy \citep[e.g.][]{2008MNRAS.387.1431D}. From equation \ref{eqn:TcSNe}, it is clear that at sufficiently high resolution this is no longer necessary. We only reach such a resolution, however, for our highest resolution runs. The issue is further complicated when we consider that the cooling time is really a product of the temperature and the local density squared. Yet, the local density -- by the time the SNe go off -- may already be somewhat lower that $n_{\rm th}$ due to the relative motions of the stars and gas, and the gas cooling over this time.

It remains to be seen whether delayed cooling and/or kinetic feedback leads to faster convergence than simply thermally dumping the energy. Such detailed issues of convergence lie beyond the scope of this present study. However, it is important for the reader to understand such subtleties when we discuss our simulation results in \S\ref{sec:discussion}.

\subsection{Metal injection and diffusion}

We implement the injection of metals from SNe following the approach taken in \cite{AGORApaper} for type-II SNe. We track the production of Oxygen and Iron from stars between $8$ and $40 \msun$, using the fitting formulae of \cite{WoosleyHeger2007},
\begin{align}
M_{\rm Fe} = 0.375 e^{-17.94/m} \; \; \; \msun\\
M_{\rm O} = 27.66 e^{-51.81/m} \; \; \; \msun
\end{align}
to obtain the total mass of each element ejected as a function of stellar mass $m$. The total metal mass is then obtained from the masses of Oxygen and Iron \citep{AsplundEtal2009} as
\begin{equation}
M_{\rm Z} = 2.09 M_{\rm O} + 1.06 M_{\rm Fe}
\end{equation}
with metallicity $Z$ defined as the fraction mass of metals out of the total mass $M_{\rm total}$ of the gas particle, $Z = M_{\rm Z}/M_{\rm total}$. The metal content is initalised, at $t = 0$, to be negligible (effectively zero).

Metals are advected between particles in the same way as all non-smoothed fluid quantities (mass, entropy, velocity) are advected in the SPHS formalism, namely a process of diffusion between particles regulated by a switch based on the spatial derivative of the velocity divergence. This switch detects if particle trajectories are going to converge \emph{in advance} and switches on diffusion of the fluid quantity (in this case $M_{\rm Z}$) only as particles approach one another. The precise expression for the diffusion of metal mass between a particle $i$ and its neighbouring particles $j = 1..N$ is as follows:
\begin{equation}
\dot{M}_{\rm{Z}, \rm{diss}, i} = \sum_j^N \frac{m_j}{\bar{\rho}_{ij}} \bar{\alpha}_{ij} v^p_{sig, ij} L_{ij} [M_{\rm{Z},i} - M_{\rm{Z},j}] J_{ij}
\end{equation}
where ${\bar{\rho}}_{ij} = [\rho_i + \rho_j]/2$ is the symmetrized density, $J_{ij} = \frac{\textbf{r}_{ij} \cdot \nabla_i W_{ij}}{\vert \textbf{r}_{ij}\vert}$ is the normalised smoothing kernel derivative, $L_{ij}$ is a pressure limiter, and $v^p_{sig, ij}$ is a positive definite form of the signal velocity \citep[for more detail see][]{2012MNRAS.422.3037R}.

\subsection{SMBH sub-grid model}

We include an SMBH at $(0,0,0)$, initially of negligible mass, modelled as a sink particle with an accretion radius of $r_{\rm acc} = 0.1$ kpc. Particles that come within this distance are removed from the simulation, and their mass is added to the mass of the black hole. To avoid any numerical effects due to the unphysical wandering of the black hole, we keep it fixed at the centre of the potential at every timestep.

We clarify here that due to the size of the accretion radius, when we refer to an `SMBH accretion rate' we are simply referring to the accretion rate through the $0.1$ kpc boundary. This is not expected to be synonomous with the actual accretion rate onto the black hole, since the gas would likely settle into an accretion disc with only the lowest angular momentum part reaching the scales where it can be directly accreted \citep[e.g.,][]{PowerEtal2011}.

In our model we do not include feedback from the SMBH since we wish to isolate here the role of SNe feedback.

\section{Results}

\subsection{SNe-driven accretion flow}\label{sec:accretionflow}

\begin{table*}
\caption{Relevant parameters for each resolution. $N$ is the total number of particles for each species, $m_{\rm gas}$ is the (constant) mass of each gas particle and $m_{\rm dm}$ gives the range of masses for the dark matter particles. $M_{\rm res}$ is the `resolvable' gas mass, defined in Section \ref{sec:cooling}. $h_{\rm min}$ refers to the minimum smoothing length of the gas over the whole simulation to 6 Gyr, while $h_{\rm min, f}$ records the minimum gas smoothing length in the filamentary inflow as the disc is forming. $\epsilon$ refers to the gravitational softening length. For all simulations, $\epsilon = h$ for the gas particles. The (constant) stellar mass for each resolution is the same as the (constant) $m_{\rm gas}$ since stars are formed by converting single gas particles and do not subsequently accrete.}
\centering
\begin{tabular}{|c|c|c|c|c|c|c|c|c|}\hline
ID & $N_{\rm gas}$ & $N_{\rm dm}$ & $m_{\rm gas}$ ($\msun$) & $m_{\rm dm}$ ($\msun$) & $M_{\rm res}$ ($\msun$) & $h_{\rm min}$ (kpc) & $h_{\rm min, f}$ (kpc) & $\epsilon_{\rm dm}$ (kpc) \\ \hline
\hline
S1 & $1.5 \times 10^5$ & $1.8 \times 10^5$ & $1 \times 10^6$ & $1 - 14 \times 10^6$  & $1.28 \times 10^8$ & $0.078$ & $0.38$ & $0.45$ \\
S2 & $7.5 \times 10^5$ & $9 \times 10^5$ & $2 \times 10^5$ & $2 - 29 \times 10^5$ & $2.56 \times 10^7$ & $0.028$ & $0.069$ & $0.2$ \\
S3 & $3.5 \times 10^6$ & $4.5 \times 10^6$ & $4 \times 10^4$ & $4 - 59 \times 10^4$ & $5.12 \times 10^6$ & $0.0044$ & $0.026$ & $0.09$ \\
\hline
\hline
\end{tabular}
\begin{flushleft}
\end{flushleft}
\label{tableics}
\end{table*}

\begin{figure}
\begin{minipage}[b]{.49\textwidth}
\centerline{\psfig{file=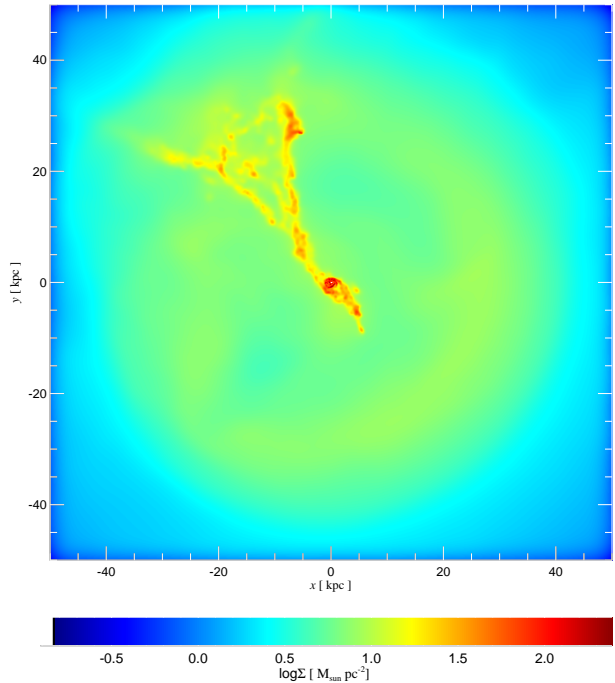,width=1.0\textwidth,angle=0}}
\end{minipage}
\caption[]{Surface density projection of the inner $50$ kpc at $t = 1.0$ Gyr. The accretion flow is seen clearly in the form of filaments funnelling gas to the centre and growing the galaxy.}
\label{fig:accretionflow}
\end{figure}

The behaviour of the simulation is much the same as that in the SPHS-96-res1 run in \cite{HobbsEtal2013}. We summarise it briefly here. The initial condition undergoes cooling and gas begins to accumulate at the centre of the computational domain, reaching the density threshold for star formation and thereby rapidly forming stars, which explode as supernovae and drive a near-spherical Sedov-like shock into the surrounding gas. This Sedov blast wave is in fact composed of many hundreds of SNe-driven bubbles, all combining to push the gas in the central regions out beyond $\sim 50$ kpc -- a `superbubble'. Gradually this hot gas cools and begins to flow back, some of it reaching the centre and driving a second starburst event at $t \simeq 1$ Gyr which is associated with the accretion flow -- we go into more detail on this later in the paper. The accretion flow, comprised of filamentary-like structures, forms a disc-like galaxy in the central $5-10$ kpc.

Figure \ref{fig:accretionflow} shows a surface density projection of the accretion flow that forms around $t \simeq 1$ Gyr. The relevant structures in this visualisation are the filamentary accretion flow, the protogalaxy forming in the central $\sim 5$ kpc, and the near-spherical hot bubble, from the 2nd main starbust event, that extends out to $\sim 50$ kpc, encompassing the accretion flow.

\subsection{Inflow rate}

\begin{figure*}
\begin{minipage}[b]{.49\textwidth}
\centerline{\psfig{file=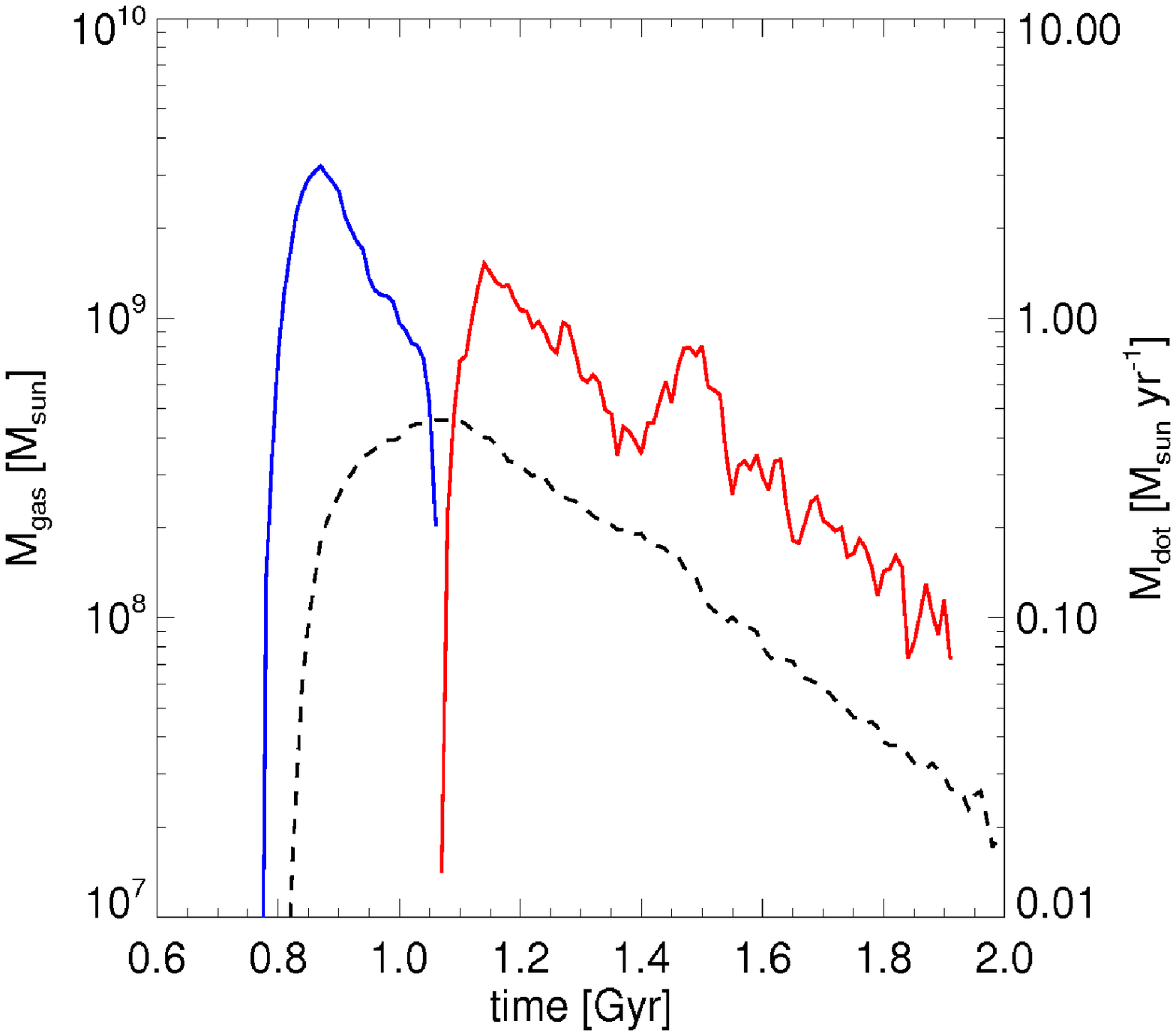,width=1.0\textwidth,angle=0}}
\end{minipage}
\begin{minipage}[b]{.49\textwidth}
\centerline{\psfig{file=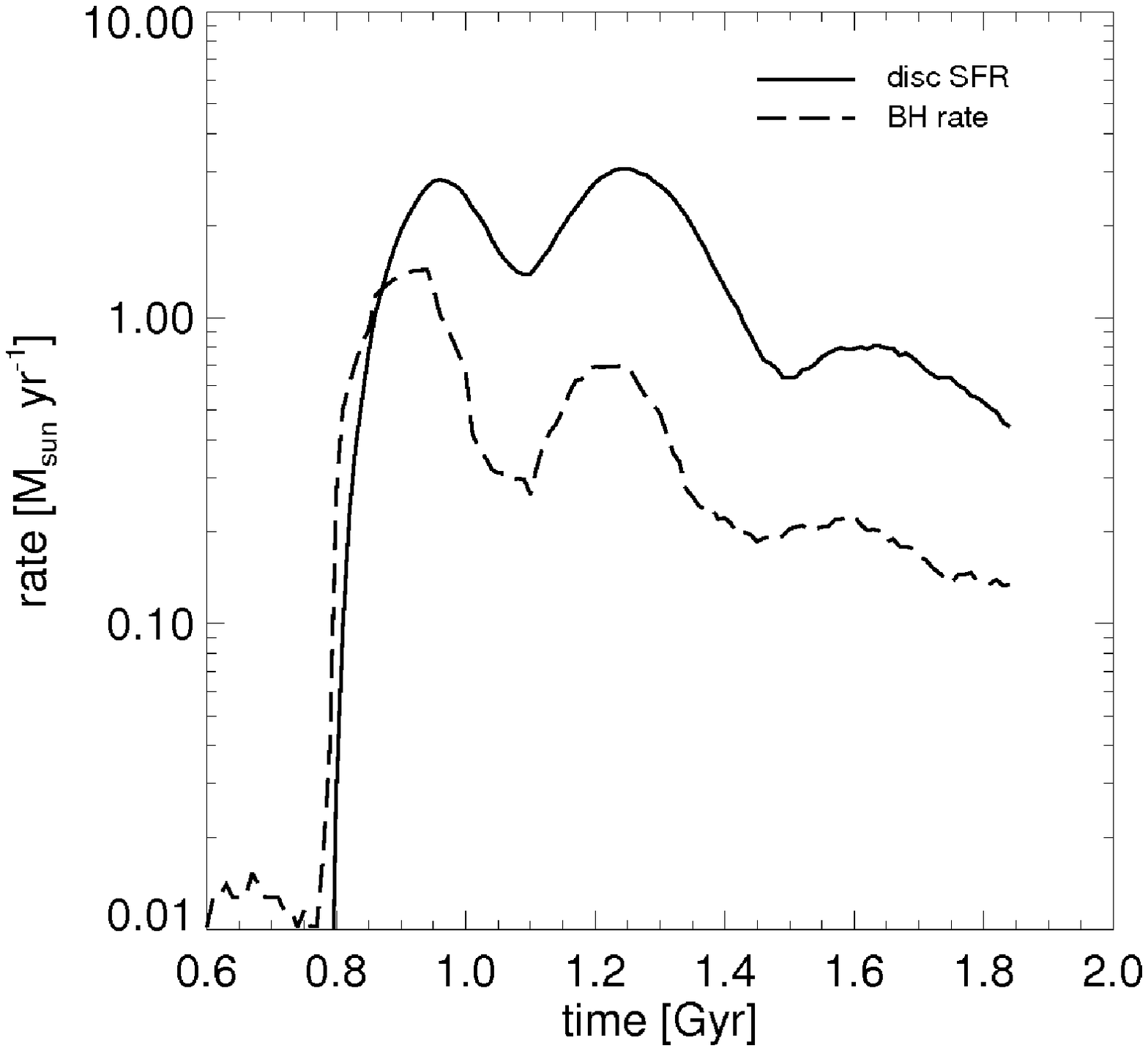,width=1.0\textwidth,angle=0}}
\end{minipage}
\caption[]{The feeding rate through the filaments (left) and the SF and $M_{\rm bh}$ rates (right) in the second main starburst event. On the left the black short-dashed line shows the amount of mass in the filaments -- left-hand axis -- while the blue and red lines refer to the rate at which gas is passing into and out of the filaments, respectively (right-hand axis). The blue is therefore the rate at which gas is condensing from the ambient medium to form the filaments, while the red is the rate at which the gas is flowing out of the filaments to the central $\sim 5$ kpc. On the right the SFR and particularly $\dot{M}_{\rm bh}$ show a very similar form to the rates in the left-hand plot, suggesting (i) the accretion flow is driving the SFR and (ii) that the accretion flow is driving gas to small radii where it can be accreted by the SMBH ($r_{\rm acc} = 0.1$ kpc in our simulation).}
\label{fig:streamsrate}
\end{figure*}

The accretion rate onto the forming galaxy through the filaments is substantial. As the left-hand plot in Figure \ref{fig:streamsrate} shows, $\dot{M} \sim 0.1-1 \msun$ yr$^{-1}$ over approx. $1$ Gyr. The condensation rate of the filaments is rapid (the blue line in the plot), at $2-3 \msun$ yr$^{-1}$. Once the filament has condensed, gas flows down through it onto the galaxy at approximately the free-fall rate. Within the galaxy itself, the SFR proceeds at a similar rate to the rate through the filaments.

\vspace{0.1in}

\noindent To leading order, the filaments deposit gas which then forms stars at the deposition rate. However, there are some interesting details. The sequence of events is as follows:

\begin{enumerate}
\item The first starburst goes off, consisting of many hundreds of different SNe bubbles each combining to produce a near-spherical collection of superbubbles that pushes the majority of the gas out to $\simgt 50$ kpc.
\item The gas gradually cools and starts to fall back down, the walls between the superbubbles starting to condense out as the gas cools.
\item One or more of these walls becomes dense enough (a factor of $\sim 10$) compared to the surrounding gas to undergo a non-linear thermal instability, leading to the rapid condensation of one or more filaments \emph{all the way along the overdense region} - from $\sim 50$ kpc down to $\simlt 1$ kpc.
\item This condensation grows an initial protogalaxy in the inner $\simlt 5$ kpc, which forms stars
\item These stars explode as supernovae, driving a second set of near-spherical bubbles outward (the protogalaxy does not have much of a disc shape by this point)
\item These outflows have a negligible effect on the already dense filaments, which begin to accrete unhindered onto the disc
\item The increase in cold gas mass in the disc drives another starburst, this time more in the form of a disc wind that is strongest above and below the plane of the galaxy. The filaments have more or less completely drained onto the disc, although some parts of the inflow have missed it and return on an eccentric orbit to add more mass to the disc.
\end{enumerate}

\subsection{Low-Z accretion}

\begin{figure}
\begin{minipage}[b]{.49\textwidth}
\centerline{\psfig{file=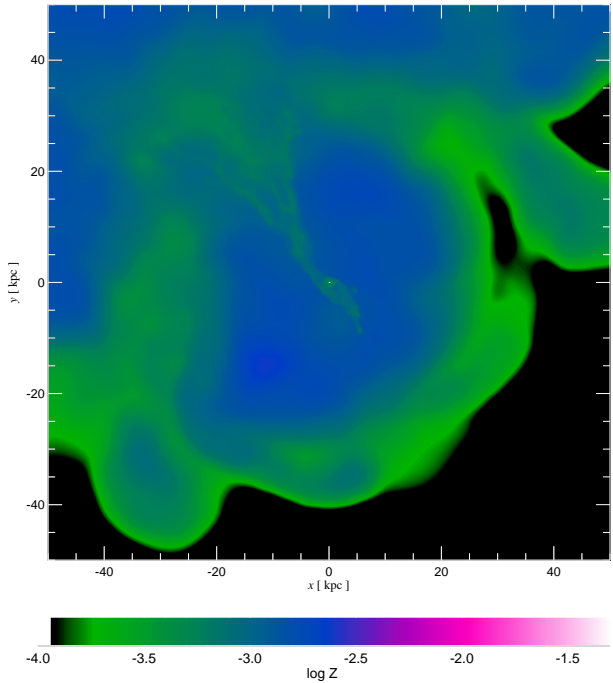,width=1.0\textwidth,angle=0}}
\end{minipage}
\caption[]{Projection of the metallicity in the inner $50$ kpc at $t = 1.1$ Gyr (refer to Figure \ref{fig:accretionflow}). The gas in the accretion flow is an order of magnitude lower in metal content than the surrounding gas at a similar radius, owing to the fact that it is being drawn from gas kicked out at an earlier epoch.}
\label{fig:metals}
\end{figure}

A common problem that arises when attempting to match observations of the Galaxy to chemical evolution models is the narrow metal abundance distribution of stars with long main-sequence lifetimes. Observational data on G-type dwarf stars \citep[e.g.,][]{Schmidt1963, WortheyEtal1996}, K-dwarfs \citep{CasusoBeckman2004}, and more recently, M-dwarfs \citep{WoolfWest2012} in the Milky Way have shown that the cumulative metallicity distribution for all of these long-lived stars is highly peaked around solar metallicity, with the tail to low $Z$ falling off much faster than predicted by `closed-box' chemical evolution models. In such a model, the production of heavy elements in stars and the subsequent enrichment of the interstellar medium (ISM) leads to a metallicity distribution that increases steadily with time, and is considerably flatter than the peaked distribution seen in observations. A solution to this is a continuous infall of metal-poor gas onto the Milky Way disc, with a total integrated infall rate of $\sim 1 \msun$ yr$^{-1}$. This dilutes the enrichment from in-situ stellar feedback.

Although our simulations constitute a relatively idealised model, we nonetheless find the right sort of behaviour for the accretion flow in terms of metal abundance. As shown in Figure \ref{fig:metals}, the streams are bringing in preferentially low-Z material to the forming galaxy. The metallicity of the accretion flow is approximately an order of magnitude smaller than the ambient medium out to $\sim 20$ kpc. We see that the initial protogalaxy, too, is under-abundant by the same factor. 

\subsection{Condensing filaments: an efficient way of funnelling gas to low $r$?}

\begin{figure}
\begin{minipage}[b]{.49\textwidth}
\centerline{\psfig{file=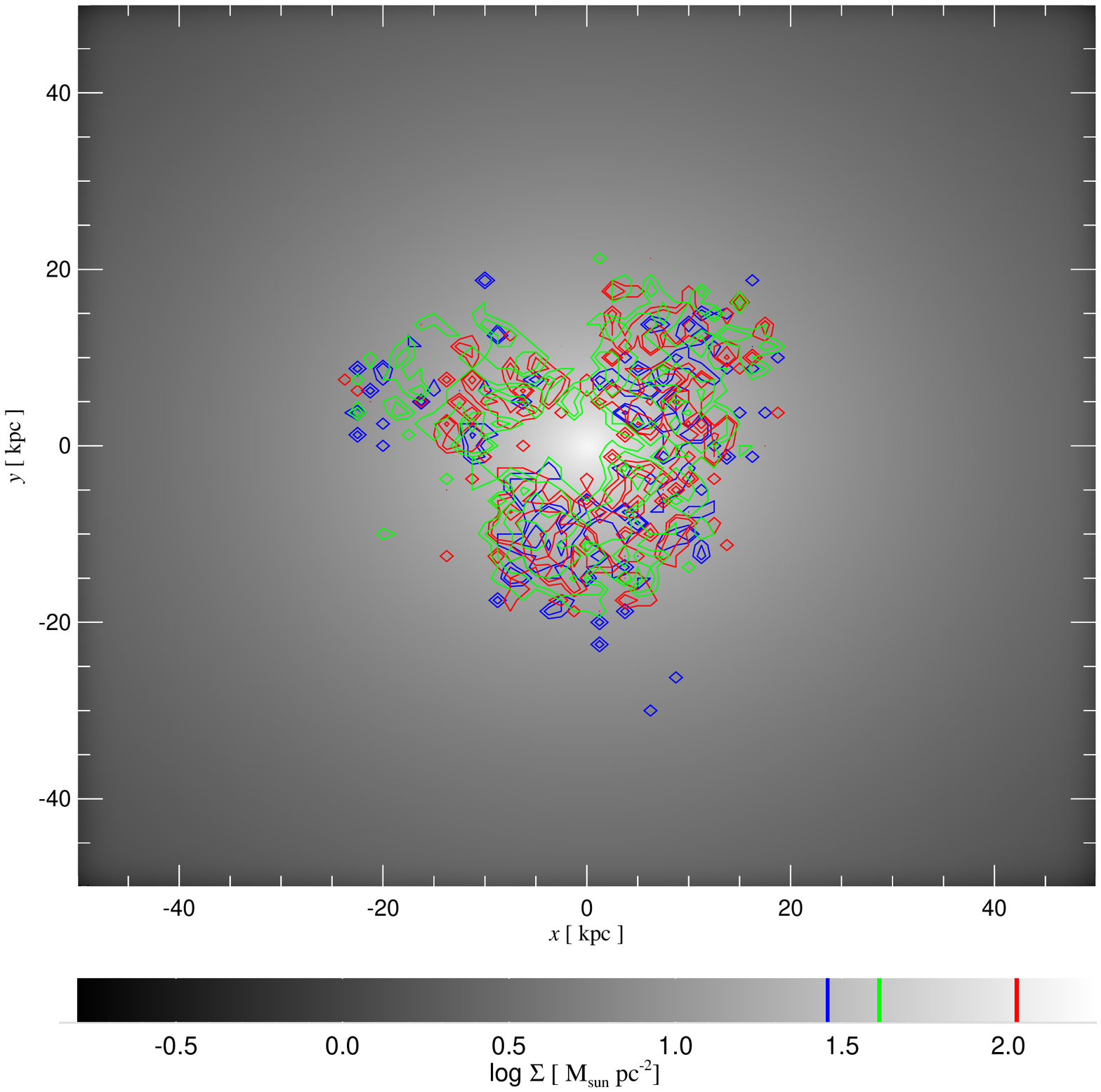,width=0.85\textwidth,angle=0}}
\end{minipage}
\begin{minipage}[b]{.49\textwidth}
\centerline{\psfig{file=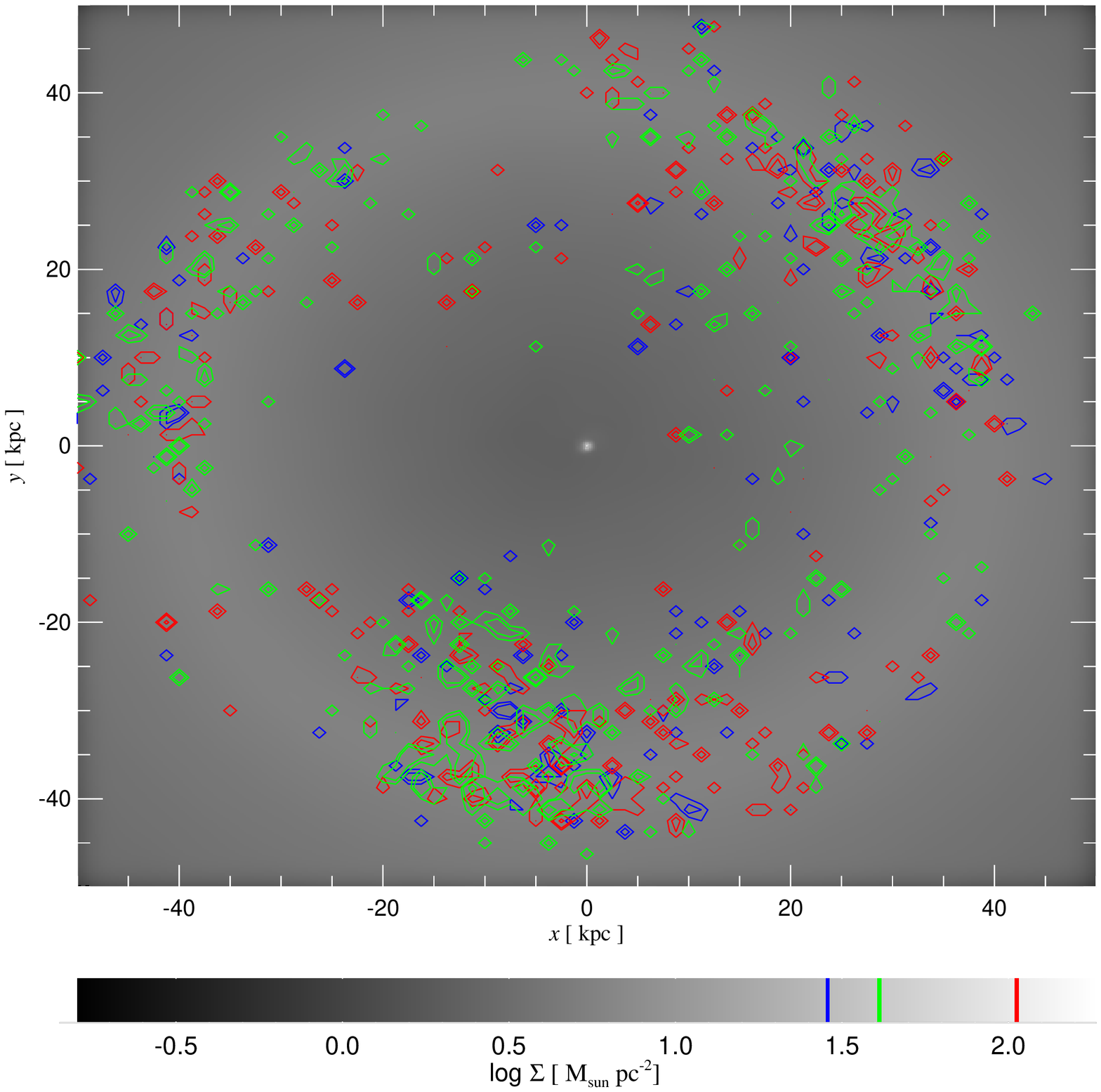,width=0.85\textwidth,angle=0}}
\end{minipage}
\begin{minipage}[b]{.49\textwidth}
\centerline{\psfig{file=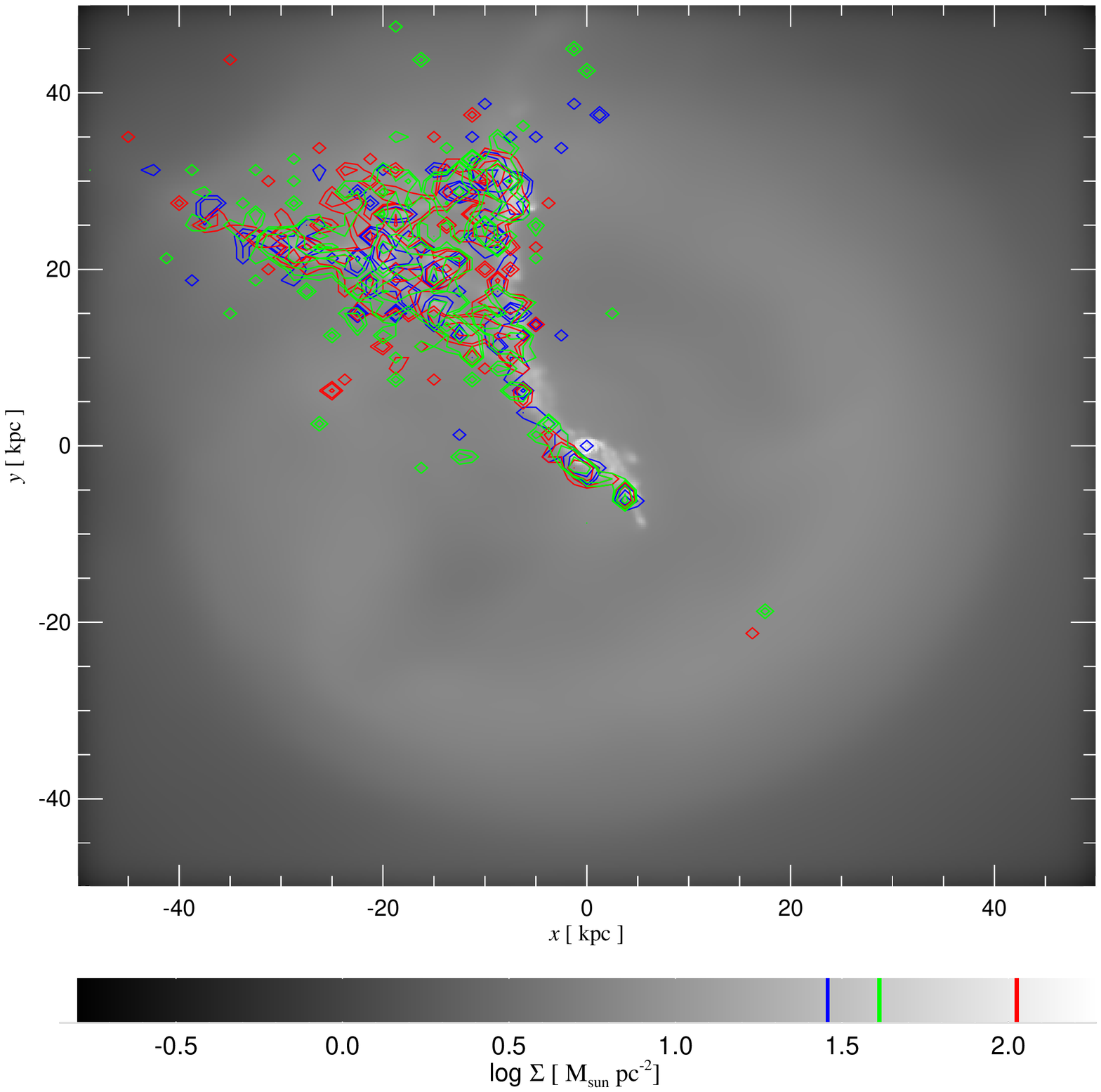,width=0.85\textwidth,angle=0}}
\end{minipage}
\caption[]{Projected surface density of gas within the inner $50$ kpc, overlaid with points identifying gas particles that end up inside a given radius range at $t = 2.0$ Gyr. Points in blue indicate gas that makes it inside $r < 1.0$ kpc, points in red for gas that makes it inside $1.0 < r < 2.0$ kpc, and in green for gas that ends up inside $2.0 < r < 5.0$ kpc. Shown here are three snapshots in time: the initial condition (top), $t = 0.2$ (middle), and $t = 1.0$ Gyr (bottom).}
\label{fig:discfeeding}
\end{figure}

We find that the filamentary accretion flow is very effective at bringing gas from large radii down to small scales. Figure \ref{fig:discfeeding} shows a surface density projection of the inner $50$ kpc at three different times. The particles that end up inside a given radius range are indicated with coloured points. A purely spherically-symmetric inflow (such as in the case of the initial velocity field) would show these particle selections in a series of shells, blue followed by red, out to only a small factor of order unity times the final radius at $t = 2.0$. In our case however, we see markedly different behaviour. There is no clear trend in $r_{\rm final}$ for the initial positions of the particles. Moreover, in some cases $r_{\rm initial} \simgt 20-40 r_{\rm final}$. We see this clearly with the initial radii for the gas particles marked in blue (those that end up within the inner $1$ kpc) which can be as high as $40-50$ kpc.

It is also clear that most of the particles making it inside this range of inner radii by $t = 2.0$ are funnelled through the main stream that forms at around $t = 0.9$. \footnote{The evolution of this feature can be seen clearly in the corresponding movie which can be found at \url{http://www.phys.ethz.ch/~ahobbs/movies.html} (movies grouped by publication and figure name).} The initial starburst re-distributes all of the marked particles into a spherical shell, largely removing any memory of their initial distributions. As the gas cools and collapses back these particles are brought in along three main curved trajectories, created by the slight asymmetries arising from the many individual feedback events that comprise the initial starbust. Two of these trajectories collide and draw the gas out into a single region along the upper-left diagonal, with a outwardly-directed radial velocity. Due to the gravity of the potential, the gas slows and comes to a stop, just as the second main starburst event occurs, which compresses the gas into one or more filaments that cool and begin to free-fall almost radially into the centre of the computational domain. The rest of the marked particles that do not feature in this strong inflow are subsequently brought in by curved trajectories created by the intersection of SN bubbles from the second main starburst.

\subsection{The angular momentum distribution}\label{sec:histograms}

\begin{figure*}
\begin{minipage}[b]{.49\textwidth}
\centerline{\psfig{file=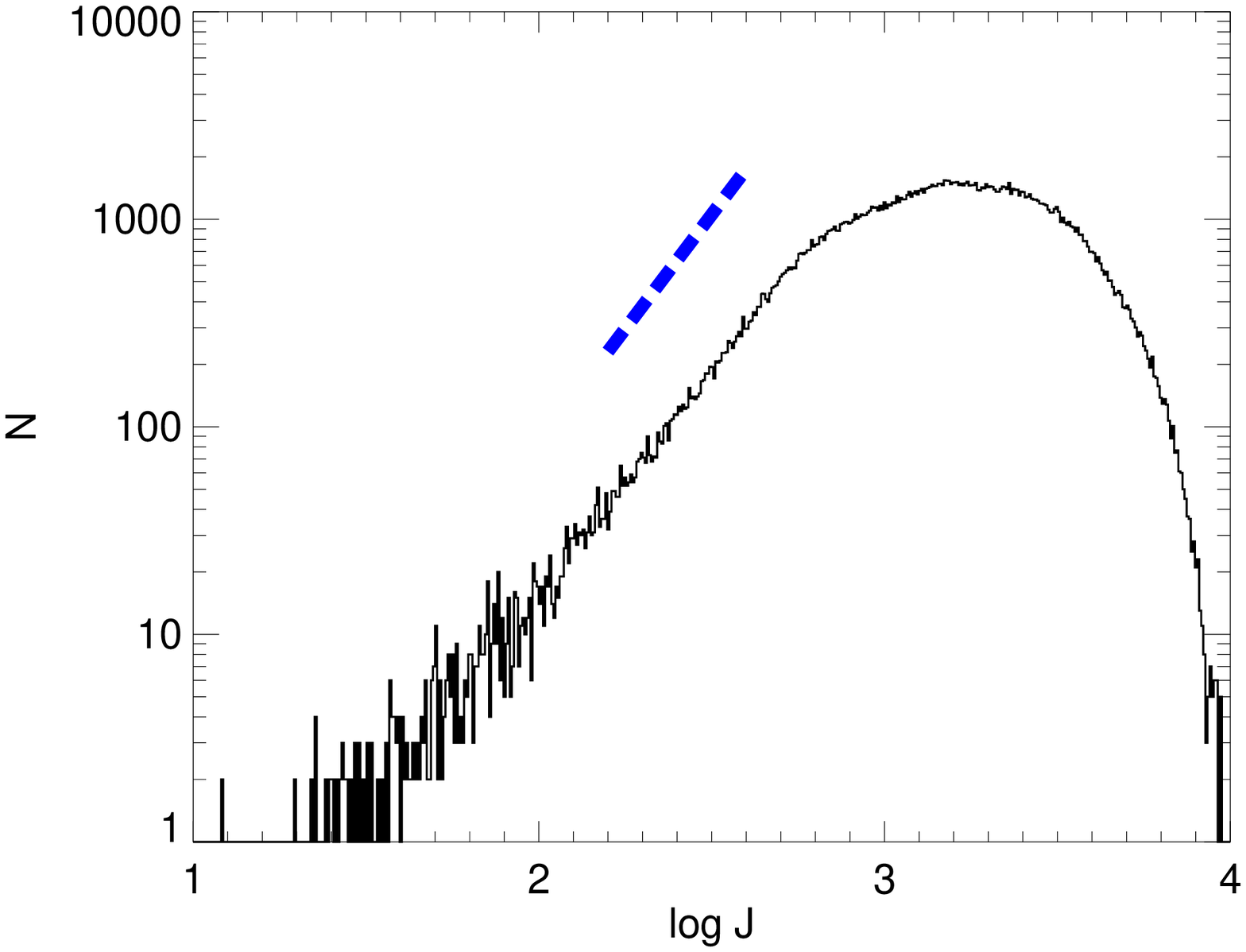,width=1.0\textwidth,angle=0}}
\end{minipage}
\begin{minipage}[b]{.49\textwidth}
\centerline{\psfig{file=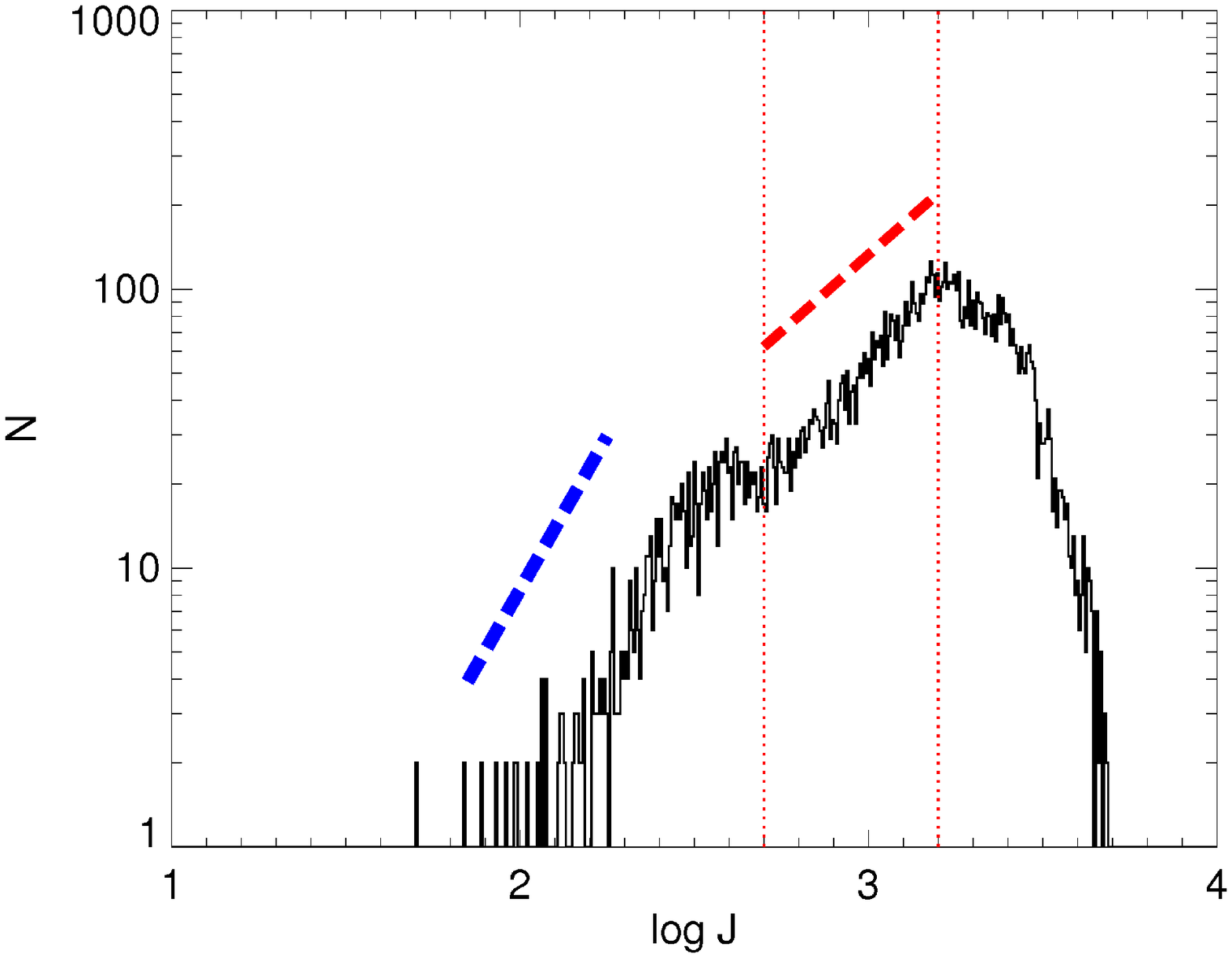,width=1.0\textwidth,angle=0}}
\end{minipage}
\caption[]{Angular momentum histograms for the all of the particles within the inner 50 kpc (left) and for only those particles that constitute the SNe-driven filaments (right). Overplotted are the analytical predictions for the slope of the low-J tail of the distribution, as derived in Section \ref{sec:lowL}: the ambient gas that follows the initial angular momentum profile (blue) and the gas condensing along a curved trajectory in the filaments (red). Note that these predictions are for the slope only, not the normalisation of the low-J tail.}
\label{fig:Lhist}
\end{figure*}

Here we study the angular momentum distribution of the accretion flow, and how it differs from the rest of the gas in the inner region. In Figure \ref{fig:Lhist} we plot angular momentum histograms for (i) all of the gas in the inner 50 kpc (left) and (ii) the gas comprising only the accretion flow (right) at $t = 1$ Gyr. The two distributions peak at approximately the same value of $\log J$, but in the left-hand case extends further to low- and high-$J$. From this plot too one can read off the angular momentum required to reach the accretion radius, since beyond this all particles are removed from the simulation -- it is approximately $J \approx 10$ in code units. The right-hand plot does not extend down this far because the contribution to the distribution from the inner $5$ kpc has been subtracted off, so as to isolate the filaments and not include the contribution from the forming galaxy.

The similarities in terms of form and peak-$J$ suggest that the filaments form from gas that is not at preferentially lower or higher angular momentum than the rest of the gas in the inner $50$ kpc. It is interesting then that one does not require a strongly low-$J$ flow to generate the streams; rather, the gas just needs to have returned to $r \simlt 50$ kpc and be cooling efficiently enough to condense out. The subsequent loss of pressure support combined with further compression by the 2nd starburst causes a rapid inflow. 

It is clear, however, that there is a difference in the tail to low-$J$ for the stream gas compared to the rest of the gas. The slope for the majority of the stream gas (delineated by the red dotted lines) is shallower than the slope expected by the velocity field from the initial condition (red vs. blue dashed lines respectively). We explore this difference in more detail, next.

\subsubsection{Characterizing the slope of the low-J tail}\label{sec:lowL}

To understand the angular momentum distribution of the filament, we can construct a simple analytical model based on insights from the simulation -- namely, that the accretion flow forms from the slightly overdense gas condensing along converging flow regions bounded by SNe bubbles. A schematic of this process is shown in Figure \ref{fig:cartoon} in the Appendix. We use the reasoning in the Figure to derive analytical predictions for the slope of the tail to low-J in the angular momentum distributions for the SNe-driven streams (as well as the rest of the gas as set by the IC). These predictions are shown in Figure \ref{fig:Lhist} alongside the actual angular momentum distributions from the simulation. Essentially, the condensing gas samples the distribution in $J$ \emph{along a curved trajectory}, giving rise to a particular exponent $\alpha = 1.1$ in the trend $M \propto J^\alpha$. We derive this assuming a circular bubble wall, presumed to be condensing out from the ambient gas as the result of a collision/convergence between one or more other SNe bubbles. While this is a relatively crude approximation to the precise geometry in the simulation (which can vary significantly) it nevertheless provides a useful insight into the mechanism as well as a good fit to the low-J tail of the angular momentum distribution.

The full derivation of this trend is outlined in the Appendix. From the right-hand panel of Figure \ref{fig:Lhist} we see that the low-J tail of the filaments is divided into two sections: one following close to (although not completely) the $M \propto j^{1.1}$ for a curved trajectory and the other following the angular momentum profile as set by the ICs, where the standard $j \propto r$ \citep{BullockEtal2001b} returns the relation $M \propto j^{2.2}$. The gas in the latter case (to the left of $\log J \sim 2.6$) simply constitutes an `atmosphere' of slightly hotter gas that surrounds the colder gas making up most of the filament. As such the slope of the distribution for these particles resembles that of the ambient gas (left-hand panel). Since there are only a very small fraction (a few hundred) of particles in this region of the plot and a correspondingly high level of noise, we take the dominant part of the low-J tail to be the section between the two red dotted lines, with the $M \propto j^{1.1}$ slope. 

\subsection{Radial variation of the cooling time: \emph{`cooling columns'}}

\begin{figure}
\begin{minipage}[b]{.49\textwidth}
\centerline{\psfig{file=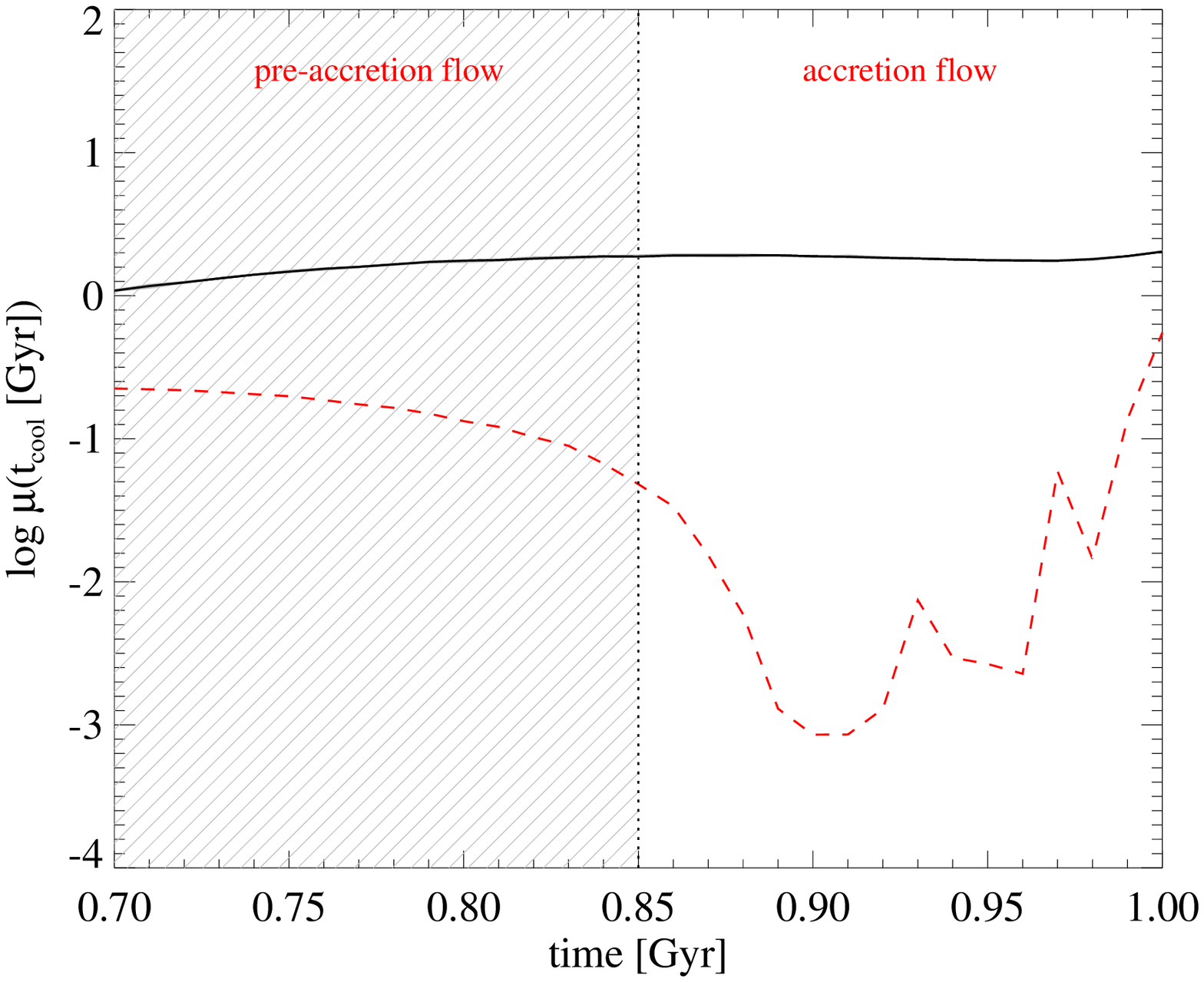,width=0.95\textwidth,angle=0}}
\end{minipage}
\begin{minipage}[b]{.49\textwidth}
\centerline{\psfig{file=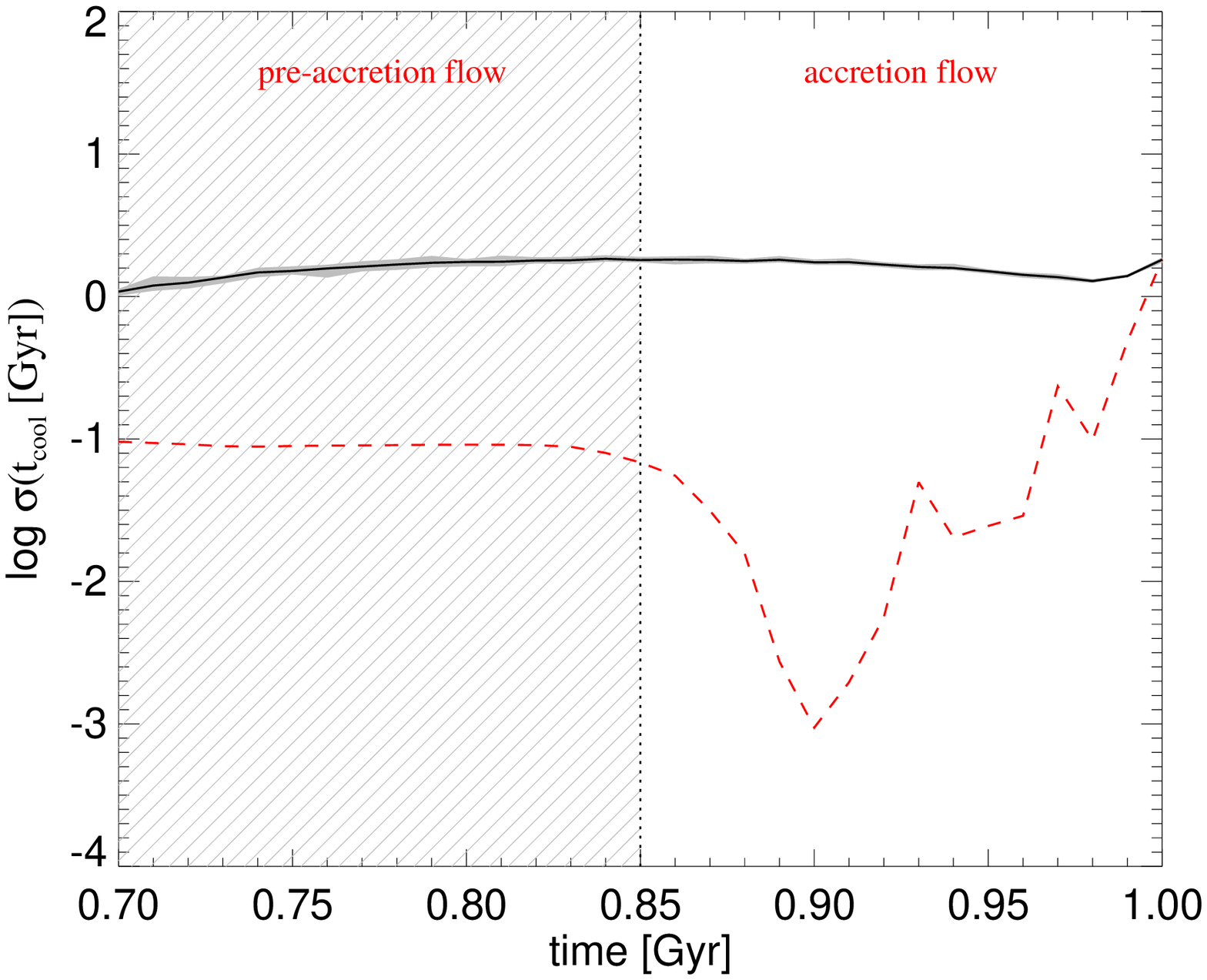,width=0.95\textwidth,angle=0}}
\end{minipage}
\begin{minipage}[b]{.49\textwidth}
\centerline{\psfig{file=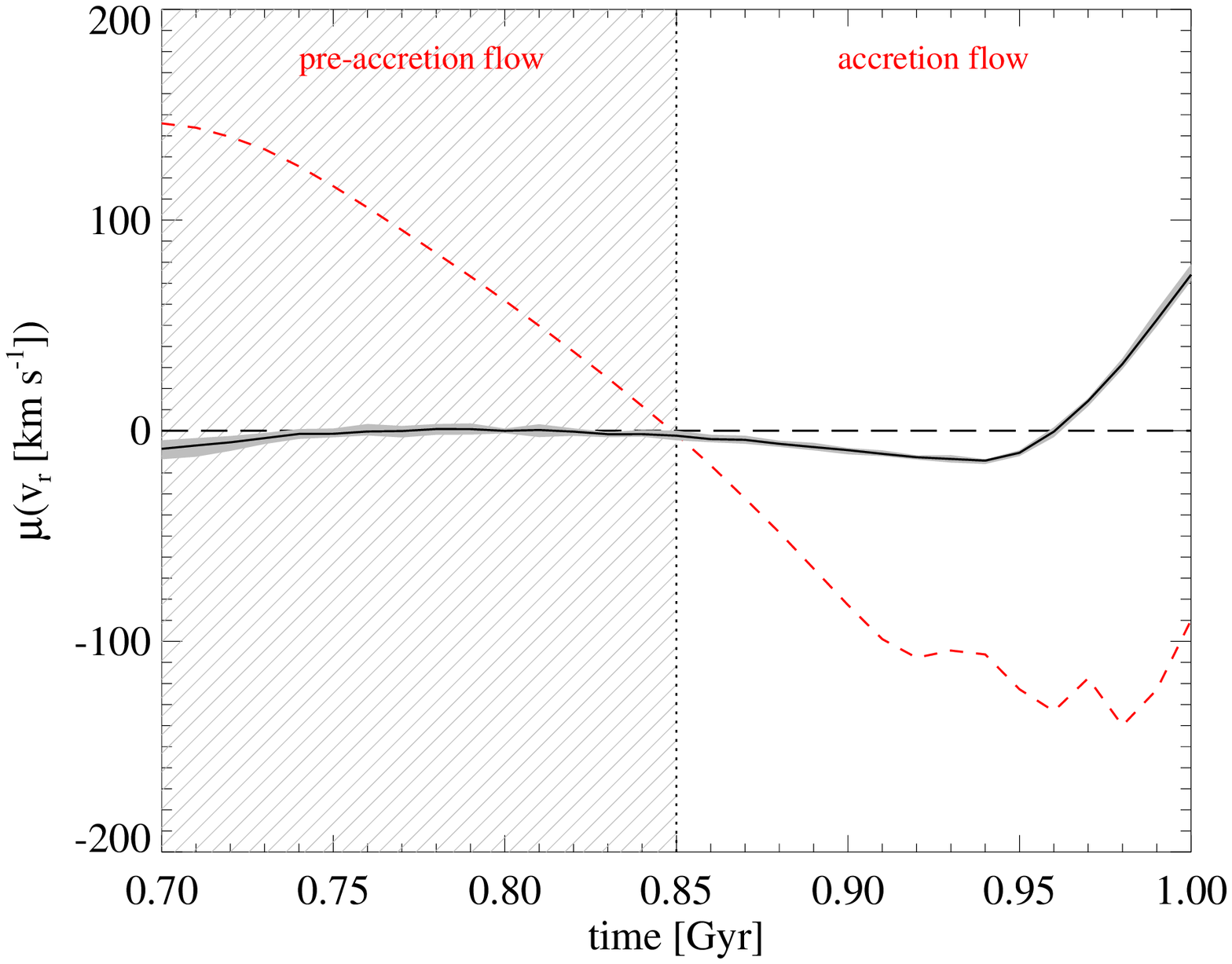,width=0.95\textwidth,angle=0}}
\end{minipage}
\caption[]{The mean (top) and standard deviation (middle) in cooling time across a $0.3$ Gyr period just before and just after the accretion flow forms, along with the mean radial velocity (bottom). Plotted is the gas that forms the accretion flow (red dashed line), and a baseline comparison (black) drawn from randomly sampling gas with the same number of particles and within the same radius range at each timestamp -- further restricted to gas that is not part of the accretion flow. We used ten random samples and averaged the values to get the black line, with the error bars on this sampling indicated by the grey shaded region above and below the line.}
\label{fig:coolingtime}
\end{figure}

In this subsection we look at the cooling time along the filament(s). From the sequence of events as laid out in Section \ref{sec:accretionflow}, we see that the accretion flow not only forms along a large range in radius but also coincides with a powerful starburst from the central $\sim 5$ kpc. This starburst is \emph{not} the result of the accretion flow feeding the disc (that comes later) but rather the gas in the central $5$ kpc condensing \emph{at the same time} as the filaments.

However, far from being a coincidence, this matching of seemingly unrelated events (separated by an order of magnitude in radius) occurs readily when the cooling time along near-radial trajectories (such as those along which the filaments form) is roughly constant. In this situation the cooling of the gas and subsequent star formation and feedback in the central few kpc occurs at approximately the same time as the cooling of the gas into filaments at larger radii. The starburst propagates outward rapidly, but the stream gas is too dense to be pushed out by the hot bubble from the starburst, which instead pushes out the low-density hot gas \citep[see, e.g.,][]{BourneEtal2014}.
   
A crucial aspect of the accretion flow and its ability to connect small- to large-scales is therefore the presence of a short cooling time along a radial column. We designate such a feature a \emph{`cooling column'}. To identify this feature in our simulation we studied the variation in the standard deviation of the cooling time over the course of the run, for both the gas that makes up the filamentary accretion flow and for the gas that doesn't (the ambient gas). To determine a statistically-significant difference in $\sigma_{\rm tcool}$, the `ambient gas' in our comparison consists of ten random samples from gas \emph{not identified as eventually belonging to the accretion flow} but each with the same number of particles as the accretion flow gas, and drawn from the same radius range as the accretion flow gas at each point in time. The values taken from these ten random samples are then averaged to produce a single value for baseline comparison. In this way we can study whether the gas that forms the accretion flow really does have a significantly lower deviation in cooling times than gas at similar radii.

The results are plotted in Figure \ref{fig:coolingtime} (middle panel). We see a clear difference in the standard deviation of the cooling time for gas that will comprise the accretion flow (red) and the ambient gas (black, with error bars indicated by shaded grey region). In particular, in the run up to the condensation of the filaments (which occurs at $t \approx 0.85$) the two standard deviations differ by more than a factor of $10$. The subsequent dip to even smaller values of $\sigma_{\rm tcool}$ (and $\mu_{\rm tcool}$) is the result of the majority of the filament gas having reached high density once the accretion flow has formed.

In addition to a low spread in $t_{\rm cool}$, the cooling column must of course also possess a low \emph{overall} value of $t_{\rm cool}$, which we confirm in the top panel of Figure \ref{fig:coolingtime} where we plot the mean value for both the filament gas and the randomly-sampled ambient gas. Also plotted in the Figure is the mean radial velocity over the same time period (bottom panel), where it is clear that the accretion flow gas starts its infall at approximately the same time, $t \approx 0.85$, as the starburst in the protogalaxy (Figure \ref{fig:streamsrate}, right-hand panel). It is also apparent that the accretion flow gas is strongly removed from the ambient gas at similar radii in terms of its radial velocity evolution.

We note that if the cooling time was not nearly constant along the length of the stream, cooling would still occur but in a more isolated fashion. For example, if only the top part of the converging flow possessed a short cooling time, only this part would condense and attempt to make its way towards the centre of the computational domain. However, there would be a great deal of hotter gas in its way, leading to it likely being stripped and mixed back in with the ambient gas before it could feed the galactic disc \citep{JoungEtal2012a}. Conversely, if only the lower part of the converging flow was to cool rapidly, it would likely make it to the disc but would be unable to bring in gas from large radii, which further imples it would be unable to bring in gas at comparatively low metallicity. Of course in general it is also clear that a converging flow that condenses only in parts would not provide as much gas as one that condenses all the way along its length.

\subsection{Resolving the accretion flow}\label{sec:resolving}

As mentioned in subsection \ref{sec:accretionflow}, the filamentary accretion flow owes to a thermal instability \citep{JoungEtal2012} that allows runaway cooling to occur. The numerical method employed in the current work has a fixed mass resolution, and although there will not be a one-to-one correspondence, we can relate a given mass scale to an approximate spatial scale by noting that regions of smaller spatial extent require higher densities for similar structures to form. In our fiducial run as presented so far, we can clearly see that the mass resolution is sufficient to resolve an accretion flow at scales of $\sim 10$ kpc. Our initial starburst is powerful, reaching $\sim 10 \msun$ yr$^{-1}$ and sending gas out to $> 100$ kpc. As such the structures that form are relatively large in scale, allowing us to resolve them. However, the radius at which particles are removed from the simulation (the accretion radius) is $0.1$ kpc, and star formation is occuring all the way down to this boundary. Supernovae feedback is therefore going on at significantly smaller (spatial) scales too, yet we see no smaller-scale copies of the accretion flow.

\begin{figure}
\begin{minipage}[b]{.49\textwidth}
\centerline{\psfig{file=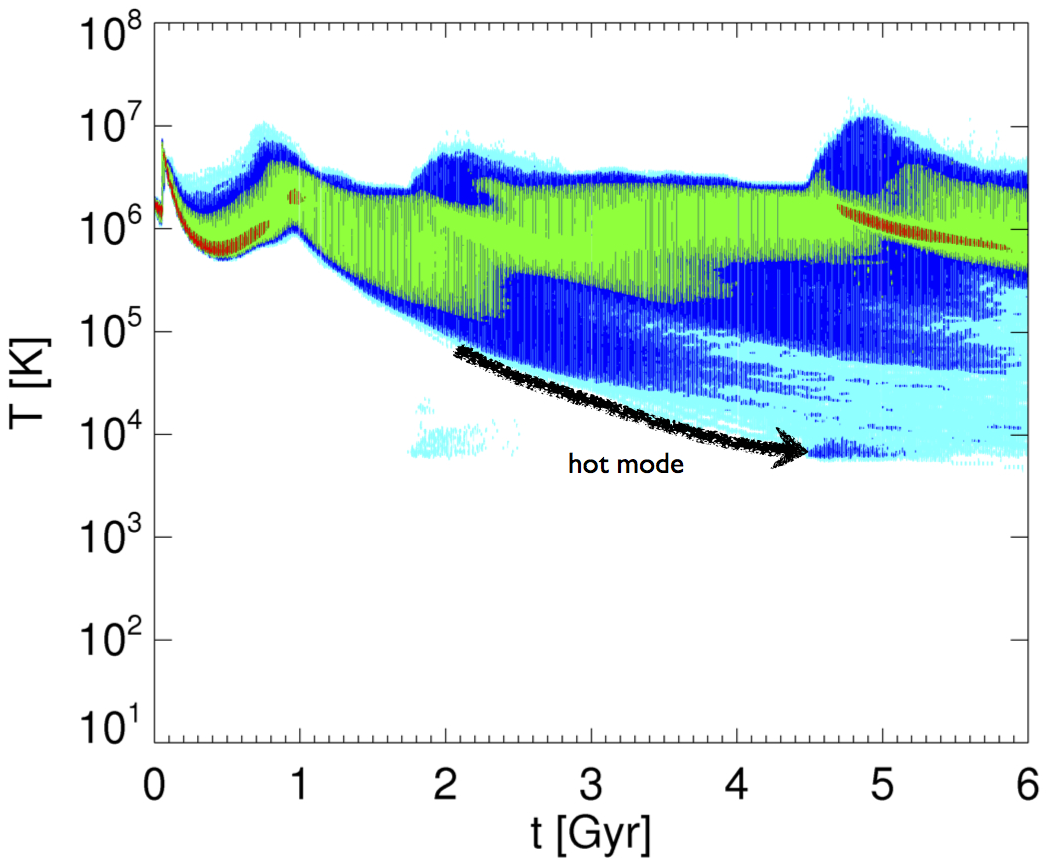,width=0.95\textwidth,angle=0}}
\end{minipage}
\begin{minipage}[b]{.49\textwidth}
\centerline{\psfig{file=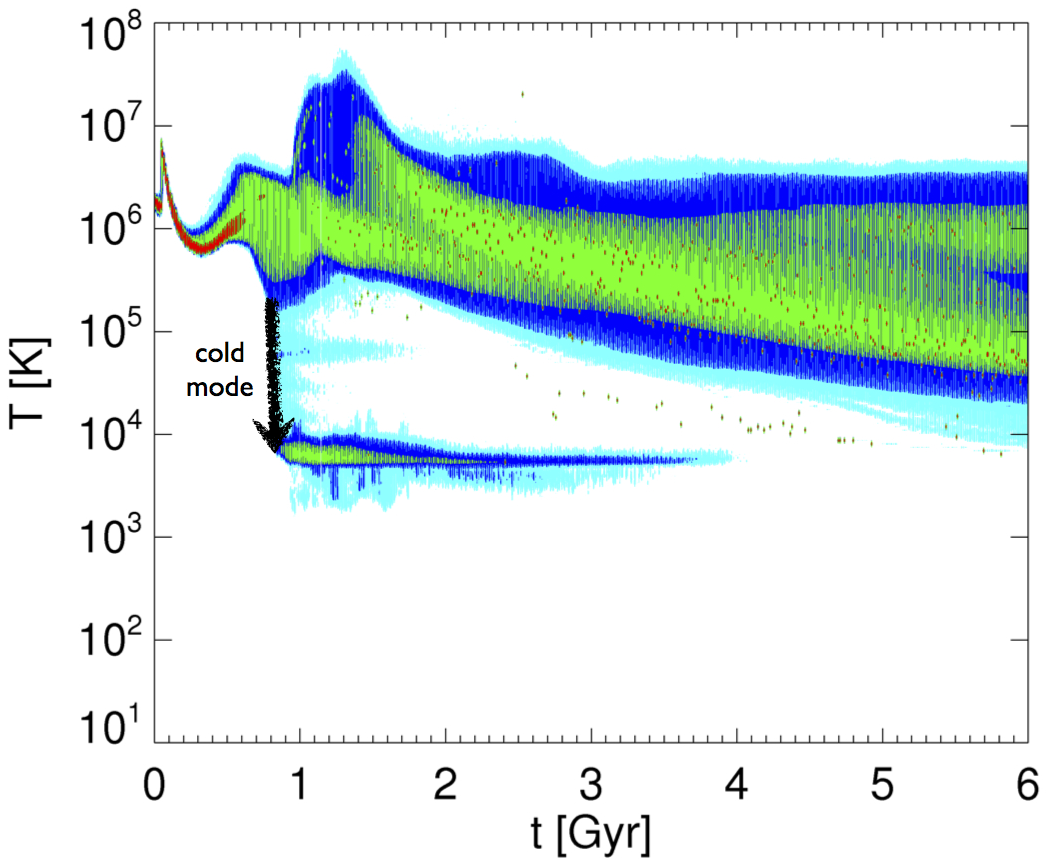,width=0.95\textwidth,angle=0}}
\end{minipage}
\begin{minipage}[b]{.49\textwidth}
\centerline{\psfig{file=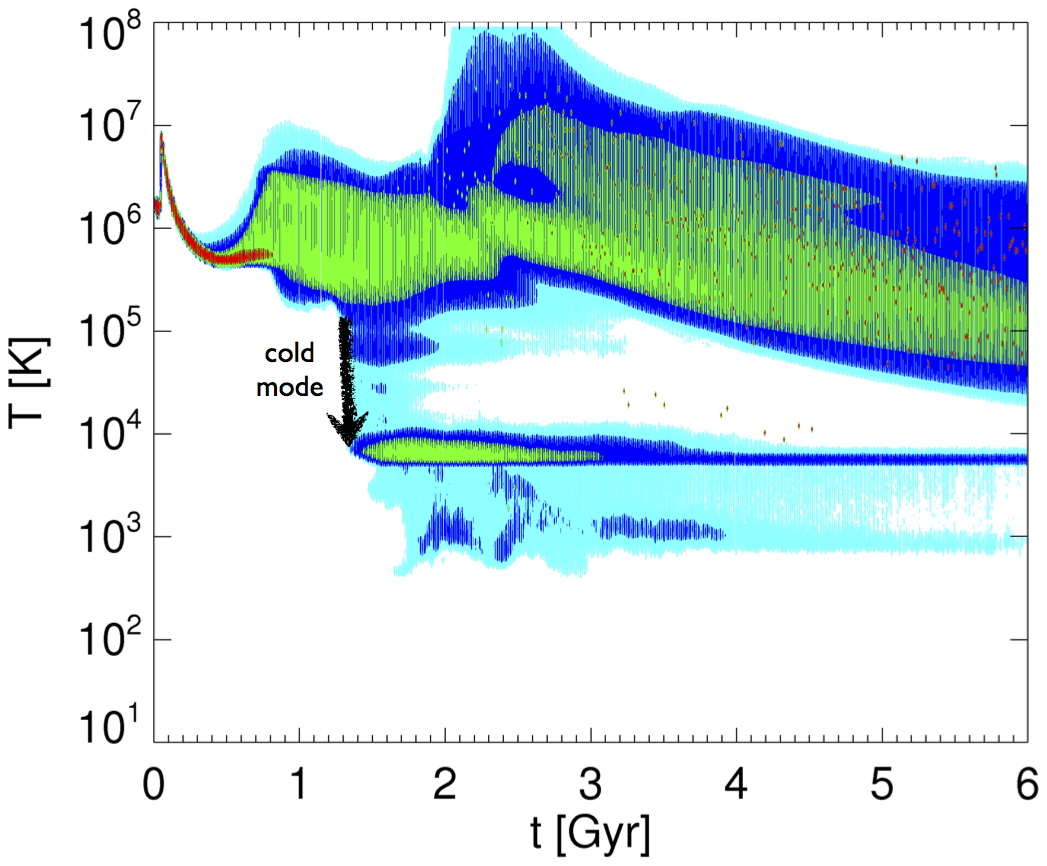,width=0.95\textwidth,angle=0}}
\end{minipage}
\caption[]{Temperature histories for gas between $10 < r < 20$ kpc at $t = 0$, tracked as the simulation evolves, for runs S1 (top), S2 (middle), and S3 (bottom). Plotted are contours of particle fraction out of the total with colours cyan $\rightarrow$ red representing levels from low $\rightarrow$ high in steps of factors of 10. The accretion flow is clearly seen in the two higher resolution simulations, and is more pronounced in S3, but is absent from S1 due to insufficient mass resolution to capture the condensation from the ambient flow. Instead, S1 undergoes a gradual cooling and loss of pressure support leading to deposition of gas into the centre. Marked on each are arrows indicating the signature of the `cold mode' sudden cooling into filaments and the `hot mode' gradual cooling of hot halo gas.}
\label{fig:resolution}
\end{figure}

We expect that such smaller-scale `ripples' leading to similar filamentary structures are present, but cannot be resolved in our fiducial simulation due to the high densities required. The difference that resolving ability makes can be seen from the different resolutions. Figure \ref{fig:resolution} shows the same run with $N_{\rm gas} = 1.5 \times 10^5$, $N_{\rm gas} = 7.5 \times 10^5$ (fiducial run), and $N_{\rm gas} = 3.5 \times 10^6$. Plotted are the temperature histories of gas in the range $10 < r < 20$ kpc in the initial snapshot. The gas particles in this initial annulus were tagged and tracked forward in time, and their temperatures recorded at each timestamp. It is clear firstly that the formation of the accretion flow (marked on the higher resolution plots as `cold mode') is very separate in its temperature evolution than the rest of the gas, breaking away quite suddenly to cool to $\sim 10^4$ K. It is also apparent that the subsequent horizontal trend that the accretion flow follows becomes more and more defined as we go up in resolution. In the lowest resolution case, the filaments do not form at all, and instead the first feature at $10^4$ K is a small fraction of the gas that has reached the centre via an `aborted' attempt to form an accretion flow, where the density convergences created a channel for gas to flow down the potential well but was not overdense enough to cool and condense out. As a result the only significant cooling occurs right at the very centre of the computational domain, giving rise to a brief lower-level star formation event that peaks below $1 \msun$ yr$^{-1}$ and drops of rapidly, reaching zero within $\approx 0.5$ Gyr. The late-time behaviour of this low resolution run is similar, and constitutes an example of `hot mode' disc growth (more on this in the next section), where the halo gas has finally cooled enough (after $\approx 4-5$ Gyrs) to become star-forming at the centre of the halo. The associated starburst is relatively weak, reaching $1 \msun$ yr$^{-1}$ only very briefly ($\simlt 0.1$ Gyr), and the corresponding protogalaxy formed from this cooling mode lacks a clear disc morphology.   

The action of the SNe feedback -- the progenitor of the density enhancement leading to thermal instability -- is also somewhat resolution dependant. As outlined in Section \ref{sec:numconv}, the SNe energy is deposited on a scale set by the smoothing length, which is naturally smaller in the higher resolution runs. Since the density at which star formation occurs is held fixed, a smaller volume for the energy dump results in a higher temperature that can be reached. Taking into consideration the typical clustering of stars formed within a main sequence time of each other, this puts our lowest resolution simulation near the peak of the cooling curve (refer to Section \ref{sec:cooling}). Due to finite time resolution, rapid cooling may lead to the thermally-dumped energy cooling away in a single timestep, i.e. before it can affect the dynamics \citep[see, e.g.,][]{2006MNRAS.373.1074S}. This constitutes a numerical/artifical cooling of the SNe energy, meaning that the feedback events are not as effective in the lowest resolution run. The thermal dump calculations for the two higher resolutions give $T_c \sim 3 \times 10^5$ and $6 \times 10^6$ respectively, corresponding to a cooling rate of more than an order of magnitude slower (for the same density) than the lowest resolution. The SNe ejecta in S2 \& S3 are therefore far less affected by numerical cooling. This helps to explain the `sea-change' non-convergent behaviour between the lowest resolution S1 and the other two, along with the explanation in terms of resolving the formation of the filaments as discussed above.

\subsection{Disc morphology}

\begin{figure}
\begin{minipage}[b]{.49\textwidth}
\centerline{\psfig{file=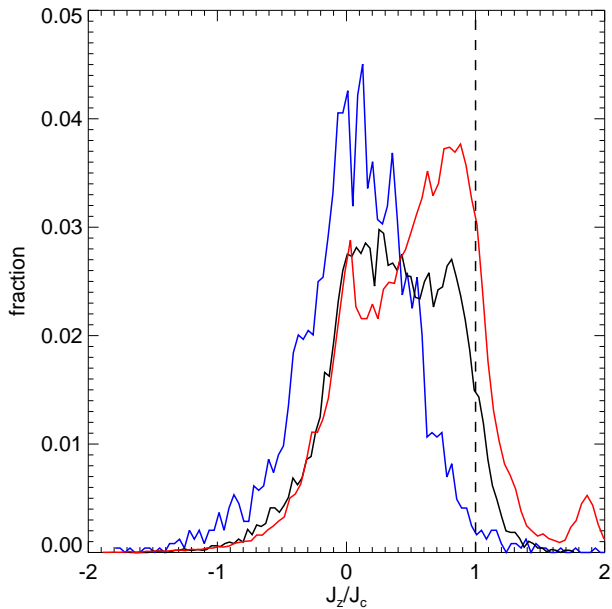,width=1.0\textwidth,angle=0}}
\end{minipage}
\caption[]{Histogram of effective disk/bulge ratio for the stellar component in each resolution, at $1$ Gyr after the main gas deposition event into the centre in each case. The mode of disk growth has a pronounced effect on the morphology of galaxy that forms, with the `hot mode' infall (low resolution, blue line) resulting in a bulge-dominated feature where $J_z/J_c$ peaks around $0$, while the cold mode infall (medium and high resolutions, black and red lines respectively) results in a more dominant disk feature at $J_z/J_c \approx 1$.}
\label{fig:diskbulge}
\end{figure}

The differences in the growth mode of the galaxy manifest strongly in its morphology. In the lowest resolution run, where the filaments do not form and the accretion is `hot mode' only, the galaxy forms with an overdominant bulge. The hot mode infall is near-spherical, without a preferential direction of rotation, and so a clear disk feature is lacking in the final galaxy. This can be seen in Figure \ref{fig:diskbulge}, where a histogram of the $J_z/J_c$ ratio is plotted for the stars (lowest resolution in blue). $J_z$ refers to the angular momentum vector along the dominant rotation axis for material inside the inner $5$ kpc, while $J_c$ is the expected angular momentum for circular orbits about this axis. As the resolution is increased and the cold mode becomes more dominant, the $J_z/J_c = 1$ signal that corresponds to a dominant galactic disk (marked on the plot as a dashed line) becomes more and more pronounced.

\subsection{Positive vs. negative feedback}

Although the SNe-driven flow is very much a `positive feedback' mode of cold gas accretion, the full situation is complex, and due to the concurrent starburst associated with the accretion flow, it partly inhibits a `hot' smooth phase of gas accretion that is simply the result of the ambient gas cooling slowly and gradually making its way down the potential well. The temperature histories in Figure \ref{fig:resolution} tell the story quite well. In particular, the lowest resolution run (top panel) cannot establish a filamentary accretion flow channel and so the dominant mode by which gas reaches the centre is this smooth phase of gradual gas cooling (marked in the figure as `hot mode'). 

This feature follows a slow trend downwards in temperature with an increasing spread as time goes on, and is also present in the two higher resolution simulations (middle and bottom panels of Figure \ref{fig:resolution}. However, in these runs it is not the means by which gas cools to $10^4$ K before we reach $6$ Gyrs. The higher resolutions runs are able to resolve the filamentary accretion flow, which appears on the temperature history plots as a near-vertical trend from high $\rightarrow$ low T (marked on the plots as `cold mode').The smooth, gradual downward cooling is still present, but the effect of the `cold mode' accretion flow -- specifically, the subsequent star formation and feedback in the central region -- is to `reset' the gradual downward cooling, sending it back up as the supernovae heat the gas. As a result, by the $6$ Gyr mark, the smooth, gradual cooling trend has not yet reached the $10^4$ K temperature level that typifies deposition of gas to the centre and allows for star formation.

We can estimate from the gradual cooling trend in the higher resolutions (simply by following it down and to the right from the time after the initial starburst) that without the filamentary accretion flow causing a second starburst the hot mode of accretion may have reached the $10^4$ K mark by $\sim 5-6$ Gyr (this is also backed up by the behaviour seen in the lowest resolution run). At this point it would most likely have had sufficiently low thermal energy compared with the gravitational potential energy to accrete onto the central $\sim 5$ kpc. However, as indeed is the case in the lowest resolution run, this accretion would not have formed a strong disc feature, as it lacks a strong preferential direction of rotation. 

The `negative' impact of the SNe-driven accretion flow could then be characterised as delaying the `hot mode' accretion by perhaps $1-2$ Gyr. At the same time, its positive impact is to entirely create an efficient `cold mode' of accretion that would not have existed otherwise.

\section{Discussion}\label{sec:discussion}

\subsection{Rapid transport of low-Z gas from large scales}

The accretion flow that we have outlined appears to be a very efficient way of transporting gas from large to small scales. In particular, it paints a markedly different picture from a neat concentric shell/ring model whereby particles closer to the centre reach it first. Instead, the streams bring in particles from a range of radii whose initial locations are seemingly uncorrelated with where they end up. The gas is drawn preferentially from a lower metallicity medium, lending support to the mechanism in terms of chemical enrichment models and corresponding observations of G-dwarfs, M-dwarf, etc. that require the galaxy to accrete low metallicity gas over its lifetime.

The filamentary flow condenses rapidly from the ambient gas, triggered by a non-linear thermal instability at the convergence of one or more SNe bubbles. Once condensed, the inflow rate is that of free-fall, with the entire filament adding its mass onto the disc in approximately a free-fall time. In our simulations, such a feeding rate lies at $\sim 3 \msun$ yr$^{-1}$ at its peak, dropping off to $0.1 \msun $yr$^{-1}$ over the course of $\sim$ a Gyr. With reference to previous work, we should make a distinction between our feeding being primarily \emph{unprocessed} gas as opposed to gas that has been processed by star formation (for example in the case of stellar winds). A simple quasi-stationary state assumption leads to an equation that predicts a very high fraction of Galactic accretion is unprocessed accretion \citep{SanchezEtal2014} which on first glance implies accretion \emph{not} caused by SNe - however in our case this distinction is not relevant since the SNe create converging flows in the unprocessed \emph{halo} gas. Our model is therefore SNe-dependant but does not have to consist of SNe-processed gas.

The presence of the filaments is a direct result of the improved treatment of hydrodynamic instabilities in SPHS as compared to `classic' SPH\footnote{we define `classic' SPH to be synonomous with the state-of-the-art equivalent to the public version of Gadget-2 when it was released \citep{Springel05}}. The latter approach produced hundreds of cold ($\sim 10^4$ K) clouds, which condensed suddenly out of uniform gas. These were spread across the majority of the hot halo, in particular anywhere where particles of significantly different entropy found themselves in the same smoothing kernel. These were demonstrated to be the result of classic SPH's inability to treat the mixing of different fluid phases \citep{HobbsEtal2013}. Without these numerical artifacts, which are ubiquitous in SPH galaxy simulations in the literature, we discovered a new set of structures that arise out of a correct treatment of mixing.\\

\subsection{High angular momentum gas $\rightarrow$ disc formation}

From Figure \ref{fig:Lhist} we can see that although the majority of the slope of the low-J tail (the section indicated by the red lines) of the filament is shallower than the corresponding slope for the rest of the gas, it is not actually at a preferentially lower absolute value of angular momentum. The peak, too, of the filaments' angular momentum distribution lies at approximately the same value as the rest of the gas in the central $50$ kpc, rather than being shifted to lower J. This is perfectly understandable in the nature of the filaments in being formed from condensing ambient gas -- there are in general no strong shocks with oppositely-directed velocities that would cancel the angular momentum, and so one would not expect a sudden decrease of J as the streams form. Indeed, a change in the slope of the low-J tail is the only thing that we would expect in this case, as derived in Section \ref{sec:lowL}. The converging flow that is the progenitor of the filaments is gradual, and merely provides the conditions for the filaments to then condense. The implications of this are that it is primarily the \emph{cooling} of the filamentary gas that drives the accretion onto the protogalaxy at the centre of the domain, rather than any direct cancellation of angular momentum. Of course, the cooling is caused by the action of SNe-outflows in compressing the gas, and so such `violent' processes play an important role.

A preferentially high value of angular momentum when forming the galactic disk is in line with what is required to form galaxies in the real Universe. Typically, simulations have suffered from galaxies that are too centrally-condensed, with an excessive bulge-to-disk ratio caused by spurious cold clouds on low-J orbits \citep{MallerDekel2002, BurkertDonghia2004, GovernatoEtal2010, HobbsEtal2013}. A mechanism that forms galaxies from a (relatively) high value of angular momentum is therefore desirable to reproduce the morphology seen in observations. From Figure \ref{fig:diskbulge} we see that as the filamentary feeding mode becomes more and more dominant (as resolution is increased) the forming galaxy has a lower ratio of bulge-to-disk.

The further implication of this picture is that our SNe-driven accretion flow will not feed the growth of the SMBH at these scales -- at least, not directly. By feeding the growth of the galaxy the flow naturally assists in getting gas closer to the SMBH but the tail to low-J as seen in Figure \ref{fig:Lhist} is still an order of magnitude away from the angular momentum at which the gas could make it inside the accretion radius, which is still $100$ pc and therefore far from the SMBH gravitational radius of influence. It is reasonable to expect, however, that similar processes could occur at smaller scales, caused either by the action of SNe closer to the centre of the galaxy or by SMBH outflows that are generated at random orientations by a precessing accretion disc and/or jet. This type of accretion flow may therefore have a role in feeding the SMBH, although further work is required to establish this.\\

\subsection{High column density features}
 
Although the model we have presented is idealised, there are several results from our simulations that already indicate a possible link to existing data. Specifically, we present reasonable quantitative agreements with data from quasar absorption lines, ram-pressure stripping of dwarfs in the MW corona, and pulsar dispersion measures, as well as general long-term SFR trends.

\begin{figure}
\begin{minipage}[b]{.49\textwidth}
\centerline{\psfig{file=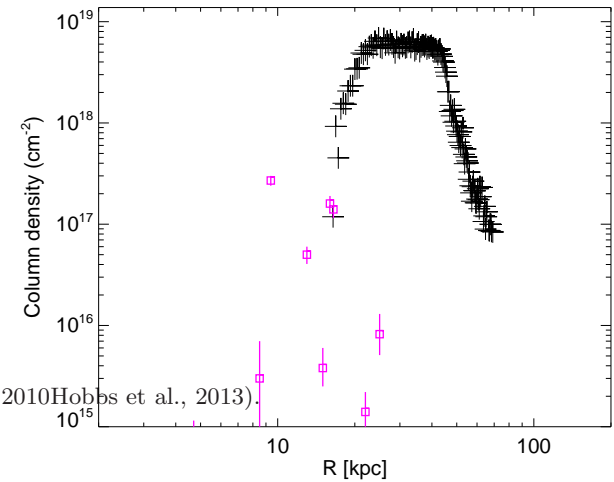,width=1.0\textwidth,angle=0}}
\end{minipage}
\caption[]{The averaged column density at $t = 0.83$ Gyr as a function of projected radius from ten random line-of-sight trajectories through the simulation domain. Overplotted are data from Mg II absorbers (magenta) as per \cite{HummelsEtal2013}.}
\label{fig:column}
\end{figure}

Recent studies of absorption along sight-lines to quasars \citep{ChenEtal2010, ProchaskaEtal2011, TumlinsonEtal2011} have yielded useful constraints on feedback models in cosmological hydrodynamical simulations \citep{HummelsEtal2013}, with the finding that, in general, models with high feedback intensity reproduce the data better. 

However, cosmological models naturally suffer from a lack of resolution in the region of interest. Figure 2 in \cite{HummelsEtal2013} shows that their highest intensity feedback model, although capable of intersecting roughly the middle of the scattered distribution of data points, has a quartile range that falls short of the highest density absorbers by $2-3$ orders of magnitude between $10-100$ kpc. In particular, the Mg II data from \cite{ChenEtal2010} indicates structures with high column density ($\sim 10^{17}-10^{18}$ atoms cm$^{-3}$) at radii of $10-20$ kpc in the corona. This discrepancy is most likely due to poor resolution in the inner parts of the halo, with the \cite{HummelsEtal2013} simulations reaching only $\sim 500$ pc for their smallest resolvable scale.

Fortunately, smaller-scale dedicated simulations such as the runs presented in the current paper can treat these inner $10-100$ kpc regions with far higher resolution. As can be seen in Table 1, our minimum resolvable scale varies from $4-80$ pc across the resolutions in the simulations as a whole, and specifically at the time of filament formation and disc growth (the crucial time for structures to condense out of the hot halo) the minimum scale is $380$ pc (lowest res), $69$ pc (medium res) and $26$ pc (highest res). The latter two runs, where the SNe-driven filaments are seen clearly, are therefore $\sim$ an order of magnitude better resolved than in \cite{HummelsEtal2013}. Interestingly, we notice that our lowest res run, where the condensation of the gas into filaments does not occur (depsite the same conditions as the higher resolutions) has a minimum resolution similar to in \cite{HummelsEtal2013}. 

We therefore ask the question, can the filamentary structures seen in our simulations reach the high column densities indicated by the Mg II absorbers? Plotted in Figure \ref{fig:column} is the averaged result of 10 random projections through the central $200$ kpc of our fiducial S2 run, at a time when the filamentary feeding mode has begun to form. The averaged column densities seen at the radii corresponding to the filaments is high, overlapping with the peak of the Mg II data at its lowest point and rising a factor of $\sim 10$ above. It would seem that our mode of feeding creates features easily dense enough to be consistent with the extreme end of the absorption data.

\begin{figure}
\begin{minipage}[b]{.49\textwidth}
\centerline{\psfig{file=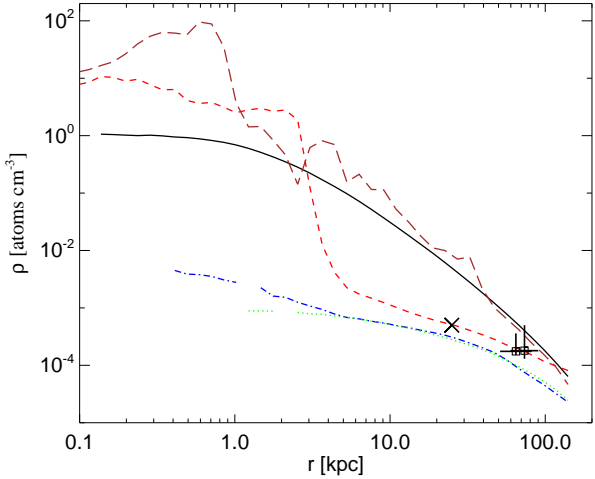,width=1.0\textwidth,angle=0}}
\end{minipage}
\caption[]{Evolution of binned density profile for fiducial run at various times: $t=0$ (black), $t=1$ Gyr (brown), $t=2$ Gyr (red), $t=4$ Gyr (blue), $t=6$ Gyr (green). Overplotted are density constrains from Sextans and Carina (black, with error bars) and the averaged density constraint from the pulsar DM measure of \cite{2010ApJ...714..320A}, set at the median radius in the range (black cross).}
\label{fig:profile}
\end{figure}

The MW corona has also been studied in terms of its ability to ram-pressure strip dwarf satellites along their orbits. \cite{GattoEtal2013} derive lower and upper bounds for the volume density of the corona at slightly larger distances, $\sim 50-90$ kpc, from the relative amounts of gas depletion (assumed to be due to ram-pressure stripping) in Sextans and Carina.

The data points and error bars from Sextans and Carina lie close to the simulation density profile at $t = 2$ Gyr, along with the median pulsar DM measure (black cross). While our model is idealised and therefore any detailed quantitive comparisons are ill-advised, the fact that the post-starburst and post-filament state of the outer halo is in agreement with the data suggests that our accretion flow process yields realistic results. Furthermore, the initial density profile (black line in Figure \ref{fig:profile}) is only a factor of $\simlt 2$ from the allowed range from Carina, and indeed is within the error bars for Sextans. This implies that the powerful starburst seen at the beginning of our simulations is not out of the question for a Milky Way type galaxy, particularly if major or even minor mergers were to be invoked. As a result the filamentary mode of galaxy build-up that is the focus of this paper seems to be allowed for by the data.\\

\subsection{Late-time accretion in spiral galaxies \& sustained SFR}

\begin{figure}
\begin{minipage}[b]{.49\textwidth}
\centerline{\psfig{file=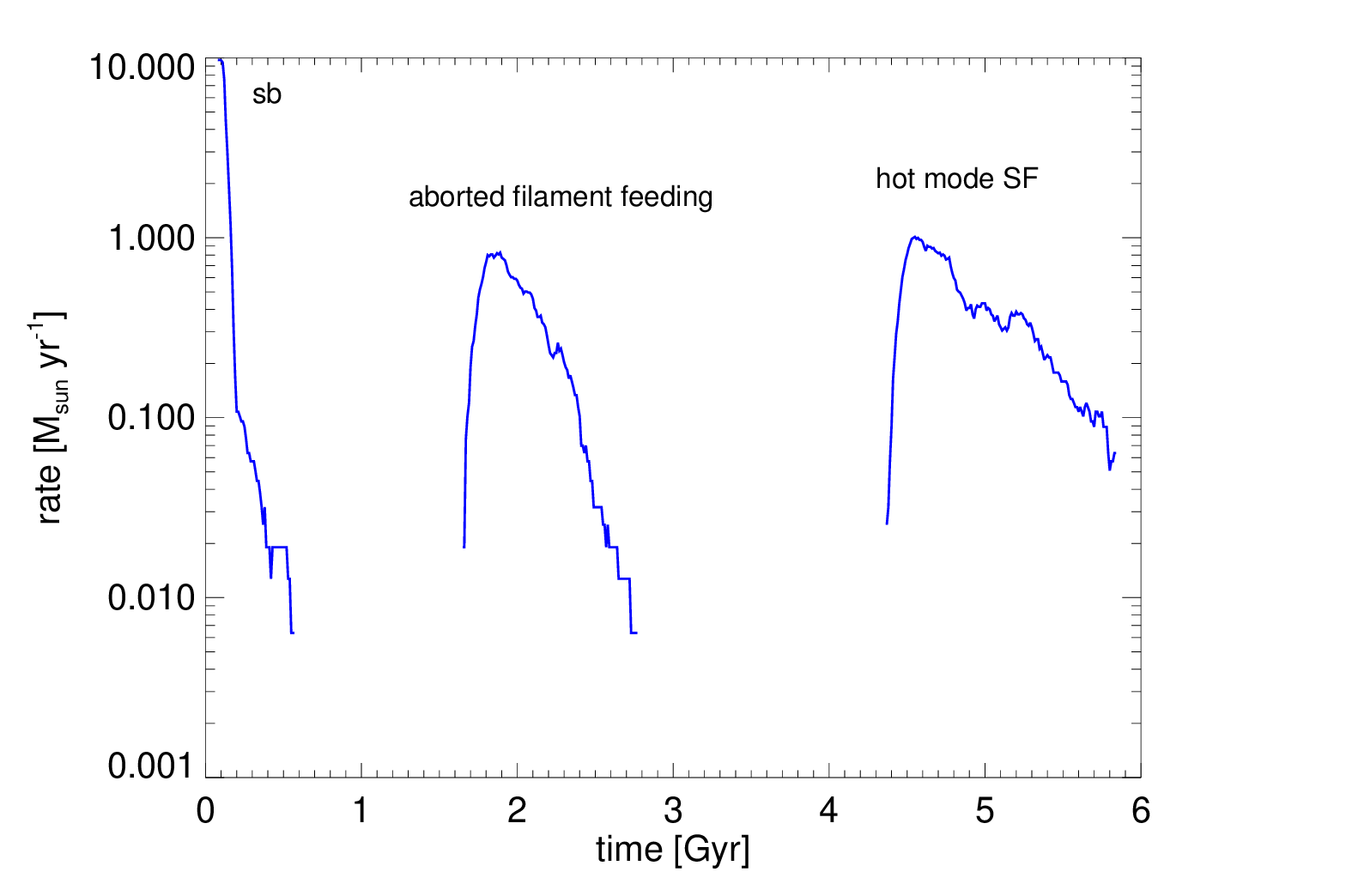,width=1.0\textwidth,angle=0}}
\end{minipage}
\begin{minipage}[b]{.49\textwidth}
\centerline{\psfig{file=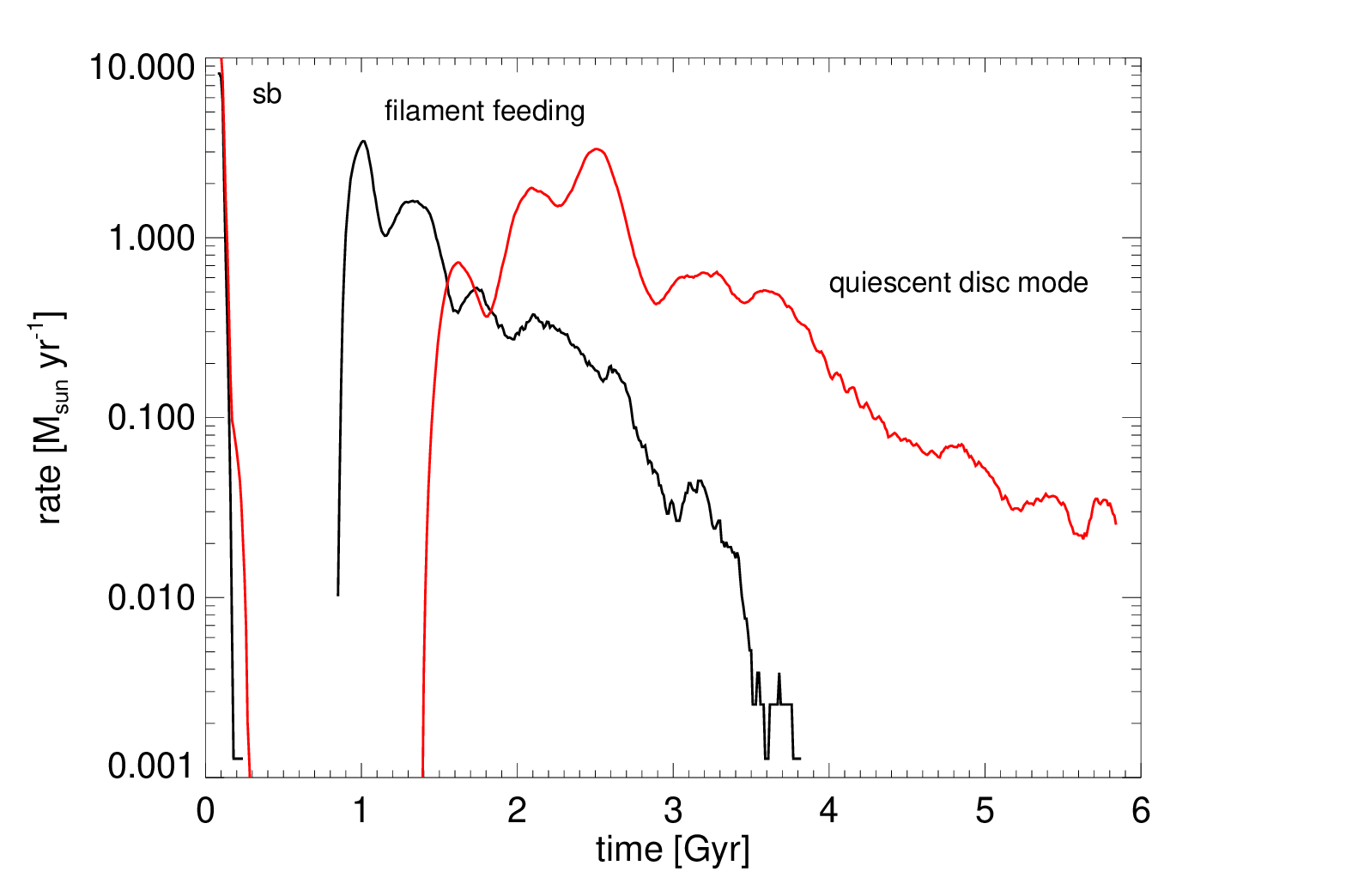,width=1.0\textwidth,angle=0}}
\end{minipage}
\caption[]{SFR to 6 Gyr for the lowest resolution (top) and the two highest resolutions (bottom). The processes responsible for each feature are marked on the plot. The lowest resolution $N_{\rm gas} = 1.5 \times 10^5$ run cannot resolve the filament formation process and therefore we see a `hot mode' channel for star formation. The higher resolutions - $N_{\rm gas} = 7.5 \times 10^5$ (black) and $N_{\rm gas} = 3.5 \times 10^6$ (red) are able to form the filamentary accretion flow and so this `cold mode' is the dominant channel for SF. The evolution of the simulation is highly chaotic in that small changes to the initial setup (such as resolution) lead to changes in the locations and rates of the initial star formation and supernovae. Subsequently, the second starburst occurs at different times for each resolution. The first starburst is marked `sb'.}
\label{fig:sfrres}
\end{figure}

We can see from Figure \ref{fig:profile} that for $t \simgt 4$ Gyr, the gas density in the halo is very low. Without a fresh source of gas, further starbursts and filamentary feeding cannot occur to late times in our simple model. This is seen in the low SFR to late times, which is shown in Figure \ref{fig:sfrres} for all three resolutions. However, we can also see from this Figure that as the resolution is increased, the late time SFR is increased by $\simgt$ an order of magnitude and decays over a longer time period. As discussed in Section \ref{sec:resolving} the higher the resolution the greater the ability of the numerical method to capture the condensation of the filaments. In general this does not mean a higher peak SFR (since this is largely set by the free-fall rate through the filaments as they initially form) but instead a greater distribution of filamentary flows across the computational domain\footnote{A movie of this process can be found at \url{http://www.phys.ethz.ch/~ahobbs/movies.html}} that lead to a more sustained SFR to 6 Gyr.

In our lowest resolution run (top panel of Figure \ref{fig:sfrres}) there is no sustained SFR. Instead, the later starbursts are short-lived. In particular, the starburst at $t \sim 5$ Gyr is not caused by our SNe-driven accretion flow due to the poor resolution; instead, it is the result of a gradual cooling of the hot corona that we term `hot mode' accretion. The fact then that increasing the resolution yields (i) a greater amount of `cold mode' filamentary accretion in the runs that can resolve the filament condensation and (ii) a completely different mode of accretion in the run that is unable to resolve filament formation is a very important result - it implies that the accretion and SFR history of late-type galaxies may undergo a `sea-change' once a threshold in resolution is crossed. This has striking implications for large-scale simulations. Moreover, the SFR attained (briefly) by the hot-mode accretion is a factor of $\sim 3$ less than the cold mode.

In the wider cosmological context, a recent paper, \cite{DiTeodoroFraternali2014}, makes the case that the late-time gas accretion from dwarf galaxies is relatively small, failing to account for the sustained SFR at late times. However, if we view such minor mergers in the context of the current work, the action of the merger, although small, could drive a starburst that would excite our SNe-driven accretion mechanism, causing gas to subsequently rain down on the disc. Events that might otherwise be dismissed as insignificant in terms of the overall accretion history might, through this process, give rise to far higher levels of accretion that \emph{are} significant.\\

\subsection{Convergence/non-convergence}

We have dealt with the issue of non-convergence in Sections \ref{sec:numconv} and \ref{sec:resolving}. Essentially, there are three aspects of the simulations which result in non-convergent behaviour, one of which is a fundamental trait of the accretion flow mechanism (forming through thermal instability) while the other two are numerical considerations in the SF and feedback strategy. As discussed in Section \ref{sec:resolving}, the fact that the filaments form via density enhancements means that as resolution is increased and higher densities are more easily reached, such a situation is able to occur more readily. The use of the polytropic cooling floor (employed to ensure that the Jeans mass is always just above the mass resolution limit) prevents the gas from collapsing to arbitrarily high densities/low temperatures, but the location of the polytrope on the $\rho-T$ diagram shifts with resolution, so that a higher resolution run can reach higher densities for a given value of $T$.

Since the SF density threshold $n_{\rm th}$ is held fixed, the above manifests as a numerical non-convergence trait in the SF strategy. Different resolutions must move along the polytrope by different amounts before reaching $n_{\rm th}$ and forming stars. For an effective temperature floor of $10^3$ K (in the code this is set as $100$ K but from the temperature history plots in Figure \ref{fig:resolution} we can see that typically the lowest $T$ reached is $10^3$ K) we require a resolution of $\sim 10$ times our highest resolution run for the gas to reach the SF density threshold without having to move along the polytrope. We anticipate that convergence could be achieved at this resolution, provided that the polytropic floor becomes `frozen' at this mass resolution, so that collapse to ever smaller scales is prevented. Such a convergence study lies beyond the scope of the present work, being challenging to run on current hardware. We will consider this in future work.

Similar to in the SF strategy, we also expect non-convergence due to the stellar feedback strategy (see Section \ref{sec:sf}). This occurs because we deposit the same energy in ever smaller volumes as the resolution is increased, leading to higher ejecta temperatures $T_{/rm dump}$ with longer cooling times. As with the SF strategy, however, we can expect to reach convergence once $T_{\rm dump}$ significantly exceeds the peak of the cooling curve. This already occurs for our highest resolution simulation.

\subsection{Similarities/differences from previous work}

Our feedback-driven filaments have similarities with the work of \cite{MarinacciEtal2011, MarascoEtal2012} in which parcels of cold metal-rich gas were sent out into the hot corona, by SNe feedback, to mix with the hotter metal-poor gas and cool it, allowing it to accrete onto the galaxy. This `ploughing' of the hot halo appears to be quite effective in sustaining disk SF at scales of the disk-corona interface. Our approach is somewhat different but still lies within the `positive feedback' category whereby SNe explosions in the disk are the direct cause of subsequent external accretion onto the disk. Our mechanism is however less reliant on the chemical makeup or temperature of the ejected gas, being entirely motivated by density enhancements through converging flows. The key difference here is therefore between supernovae-driven \emph{cooling} and supernovae-driven \emph{convergence}. Once the converging flow is created, it will of course undergo a runaway non-linear thermal instability and cool rapidly, allowing the gas to accrete onto the disk at the free-fall rate. But it does not cool from the mixing of ejected gas in our case.

\cite{JoungEtal2012, JoungEtal2012a} modelled the disruption of dense clouds in a stratified hot halo, in dedicated adaptive mesh refinement (AMR) simulations. There are a number of results from this paper that agree with our findings. First, runaway cooling only occurs in sufficient overdensities of a factor of $\sim 10$ \citep[see also][]{HobbsEtal2013}. Second, the surrounding hot halo medium was found to not cool along with the overdensity - the dense region was decoupled in terms of its thermodynamic properties from the rest of the gas. Similar behaviour was seen in our runs for the filamentary gas once it formed. From the top panel of Figure \ref{fig:coolingtime} we see that the mean $t_{\rm cool}$ of the surrounding ambient gas remains relatively flat as we transition from pre-accretion flow to accretion flow. In Figure \ref{fig:resolution} the filamentary overdensity is clearly separated in its temperature properties from the rest of the gas in the $10-20$ kpc shell.

However, an important difference in our work from \cite{JoungEtal2012a} is the ability of the cold gas to reach the galaxy without being disrupted. This is due to the formation of a dense accretion \emph{channel} through which gas can flow while remaining `protected'.\\

\subsection{Superbubbles as a general accretion mechanism}

The requirement for our accretion flow is convergences between superbubbles. However, it is only due to our relatively simple setup that these superbubbles are created only by supernovae. We fully expect that, were AGN feeding and feedback both included and resolved to a reasonable degree, powerful outflows from the SMBH would, if sufficiently numerous and randomised in angle, create similar flows. We therefore view our `superbubble-driven accretion flow' as a general mechanism for accretion that may occur wherever and whenever multiple superbubbles are generated.

\section{Conclusions}\label{sec:conclusion}

The action of SNe feedback is typically viewed as hindering accretion to smaller scales. We have demonstrated in this paper that this is not always the case. Multiple asymmetric SNe events can, if sufficiently powerful, drive converging flows in the hot halo that reach sufficient density relative to the surrounding gas to undergo a runaway thermal instability. The result of this is a high accretion rate from scales as far out as $\sim 50$ kpc and an equivalent star formation rate in the galaxy itself.

There is no real evidence of cosmological infall onto late-type galaxy discs at $z < 1$, yet, the majority of the stars were formed after this time \citep{AumerBinney2009, Fraternali2014}. In order to explain the continued high rate of star formation one must appeal to gas replenishment. We have presented an entirely new mode of gas accretion onto late-type galaxies, through a channel of `positive feedback' caused by supernovae. The SNe-driven accretion flow is (i) highly efficient at bring gas across more than an order of magnitude in scale (large to small) (ii) comprised preferentially of low-metallicity gas (iii) able to provide the forming galaxy with a gas replenishment rate of $0.1-1 \msun$ yr$^{-1}$. We have characterised the accretion flow in terms of its angular momentum distribution and derived a simple analytical explanation for the low-J tail. We have further identified a potential observational `smoking gun' for the accretion flow, in the form of `cooling columns' whereby a radial column of $\sim 1-50$ kpc has a peak cooling time \emph{and} spread of $\sim$ an order of magnitude lower than the surrounding gas.  

The accretion flow can be characterized as a `cold mode' (in the form of filaments) embedded within a `hot mode' (in the form of a rarefied atmosphere). It is expected that such SNe-driven accretion flows will be present in many other galaxy formation simulations that possess sufficient resolution, and will contribute significantly to the inflow rate onto the galaxy and to even smaller scales. In this paper we have used a simple model to motivate a characterization and analysis of how these types of flows feed smaller scales, and their likely properties in terms of inflow rate and metallicity bias.

The physical extent of galactic outflows observed in the real Universe through Lyman-$\alpha$ absorption is easily large enough to account for superbubbles of the size seen in our simulations, as well as the high velocities of the outflowing gas \citep[e.g.,][]{ErbEtal2012}. Moreover, such outflows often appear long before galaxies develop extended gaseous or stellar disks \citep{Erb2013}, suggesting that their role at least in part as `galaxy builders' may be significant.

While in this paper we have dealt with a SNe-driven accretion flow on scales of $\sim 1-50$ kpc, we note that such processes are likely to occur also on smaller scales. Indeed, anywhere there are relatively strong, repeated outflows, with a degree of asymmetry (to allow for the formation of structure along the outflow boundaries) combined with efficient gas cooling, we expect that accretion flows of this nature can occur. A notable example of this could be outflows from the SMBH, which have been shown to be both (i) powerful enough to send gas to large radii \citep{KingEtal2011} and (ii) stochastic in terms of orientation due to a precessing accretion disc and jet angle \citep{KingEtal2008}.

A key point to take from our work is that the resolution of a simulation can affect how gas gets into galaxies. Late-time accretion in cosmological simulations may proceed by a completley different mode simply because the resolution is too low to capture the correct mode. By extension, this means that capturing the correct star formation history and therefore the physics of galaxy growth may not be possible below a given resolution threshold.

We acknowledge that our model is idealised, and captures only a small part of the galaxy formation process. Future work will examine the role that this mode of gas replenishment plays in the larger, cosmological context, with high-resolution cosmological simulations performed with SPHS. Using these, we will attempt to determine the relative importance of this mode of feeding compared to other sources.

\section{Acknowledgments}

\bibliographystyle{mnras}

\bibliography{references}

\begin{thebibliography}{74}
\expandafter\ifx\csname natexlab\endcsname\relax\def\natexlab#1{#1}\fi

\bibitem[{Abadi} et~al.(2003){Abadi}, {Navarro}, {Steinmetz} \&
  {Eke}]{AbadiEtal2003}
{Abadi} M.~G., {Navarro} J.~F., {Steinmetz} M., {Eke} V.~R., 2003, \apj, 597,
  21

\bibitem[{Anderson} \& {Bregman}(2010)]{2010ApJ...714..320A}
{Anderson} M.~E., {Bregman} J.~N., 2010, \apj, 714, 320

\bibitem[{Asplund} et~al.(2009){Asplund}, {Grevesse}, {Sauval} \&
  {Scott}]{AsplundEtal2009}
{Asplund} M., {Grevesse} N., {Sauval} A.~J., {Scott} P., 2009, \araa, 47, 481

\bibitem[{Aumer} \& {Binney}(2009)]{AumerBinney2009}
{Aumer} M., {Binney} J.~J., 2009, \mnras, 397, 1286

\bibitem[{Bate} \& {Burkert}(1997)]{BateBurkert1997}
{Bate} M.~R., {Burkert} A., 1997, \mnras, 288, 1060

\bibitem[{Binney} et~al.(2009){Binney}, {Nipoti} \&
  {Fraternali}]{BinneyEtal2009}
{Binney} J., {Nipoti} C., {Fraternali} F., 2009, \mnras, 397, 1804

\bibitem[{Bourne} et~al.(2014){Bourne}, {Nayakshin} \& {Hobbs}]{BourneEtal2014}
{Bourne} M.~A., {Nayakshin} S., {Hobbs} A., 2014, \mnras, 441, 3055

\bibitem[{Bullock} et~al.(2001){Bullock}, {Dekel}, {Kolatt}
  et~al.]{BullockEtal2001b}
{Bullock} J.~S., {Dekel} A., {Kolatt} T.~S., et~al., 2001, \apj, 555, 240

\bibitem[{Burkert} \& {D'Onghia}(2004)]{BurkertDonghia2004}
{Burkert} A.~M., {D'Onghia} E., 2004, in { Penetrating Bars Through Masks of
  Cosmic Dust\/}, edited by D.~L. {Block}, I.~{Puerari}, K.~C. {Freeman},
  R.~{Groess}, E.~K. {Block}, vol. 319 of { Astrophysics and Space Science
  Library\/},  341

\bibitem[{Casuso} \& {Beckman}(2004)]{CasusoBeckman2004}
{Casuso} E., {Beckman} J.~E., 2004, \aap, 419, 181

\bibitem[{Cecil} et~al.(2002){Cecil}, {Bland-Hawthorn} \&
  {Veilleux}]{CecilEtal2002}
{Cecil} G., {Bland-Hawthorn} J., {Veilleux} S., 2002, \apj, 576, 745

\bibitem[{Chen} et~al.(2010){Chen}, {Helsby}, {Gauthier}, {Shectman},
  {Thompson} \& {Tinker}]{ChenEtal2010}
{Chen} H.-W., {Helsby} J.~E., {Gauthier} J.-R., {Shectman} S.~A., {Thompson}
  I.~B., {Tinker} J.~L., 2010, \apj, 714, 1521

\bibitem[{Cole} et~al.(2011){Cole}, {Dehnen} \& {Wilkinson}]{ColeEtal2011}
{Cole} D.~R., {Dehnen} W., {Wilkinson} M.~I., 2011, \mnras, 416, 1118

\bibitem[{Dalla Vecchia} \& {Schaye}(2008)]{2008MNRAS.387.1431D}
{Dalla Vecchia} C., {Schaye} J., 2008, \mnras, 387, 1431

\bibitem[{Dehnen} \& {Aly}(2012)]{DehnenAly2012}
{Dehnen} W., {Aly} H., 2012, ArXiv e-prints

\bibitem[{Dehnen} \& {McLaughlin}(2005)]{DehnenMcLaughlin2005}
{Dehnen} W., {McLaughlin} D.~E., 2005, \mnras, 363, 1057

\bibitem[{Dekel} et~al.(2009){Dekel}, {Birnboim}, {Engel} et~al.]{DekelEtal09}
{Dekel} A., {Birnboim} Y., {Engel} G., et~al., 2009, \nat, 457, 451

\bibitem[{Di Teodoro} \& {Fraternali}(2014)]{DiTeodoroFraternali2014}
{Di Teodoro} E., {Fraternali} F., 2014, ArXiv e-prints

\bibitem[{Doyle} et~al.(2005){Doyle}, {Drinkwater}, {Rohde} \&
  1]{DoyleEtal2005}
{Doyle} M.~T., {Drinkwater} M.~J., {Rohde} D.~J., 1, 2005, 361, 34

\bibitem[{Erb}(2013)]{Erb2013}
{Erb} D., 2013, in { American Astronomical Society Meeting Abstracts 221\/},
  vol. 221 of { American Astronomical Society Meeting Abstracts\/},  428.01

\bibitem[{Erb} et~al.(2012){Erb}, {Quider}, {Henry} \& {Martin}]{ErbEtal2012}
{Erb} D.~K., {Quider} A.~M., {Henry} A.~L., {Martin} C.~L., 2012, \apj, 759, 26

\bibitem[{Fern{\'a}ndez} et~al.(2012){Fern{\'a}ndez}, {Joung} \&
  {Putman}]{FernandezEtal2012}
{Fern{\'a}ndez} X., {Joung} M.~R., {Putman} M.~E., 2012, \apj, 749, 181

\bibitem[{Forman} et~al.(1985){Forman}, {Jones} \& {Tucker}]{FormanEtal1985}
{Forman} W., {Jones} C., {Tucker} W., 1985, \apj, 293, 102

\bibitem[{Fraternali}(2014)]{Fraternali2014}
{Fraternali} F., 2014, in { IAU Symposium\/}, edited by S.~{Feltzing},
  G.~{Zhao}, N.~A. {Walton}, P.~{Whitelock}, vol. 298 of { IAU Symposium\/},
  228--239

\bibitem[{Gatto} et~al.(2013){Gatto}, {Fraternali}, {Read}, {Marinacci}, {Lux}
  \& {Walch}]{GattoEtal2013}
{Gatto} A., {Fraternali} F., {Read} J.~I., {Marinacci} F., {Lux} H., {Walch}
  S., 2013, \mnras, 433, 2749

\bibitem[{Governato} et~al.(2010){Governato}, {Brook}, {Mayer}
  et~al.]{GovernatoEtal2010}
{Governato} F., {Brook} C., {Mayer} L., et~al., 2010, \nat, 463, 203

\bibitem[{Governato} et~al.(2004){Governato}, {Mayer}, {Wadsley}
  et~al.]{GovernatoEtal2004}
{Governato} F., {Mayer} L., {Wadsley} J., et~al., 2004, \apj, 607, 688

\bibitem[{Hobbs} et~al.(2013){Hobbs}, {Read}, {Power} \& {Cole}]{HobbsEtal2013}
{Hobbs} A., {Read} J., {Power} C., {Cole} D., 2013, \mnras, 434, 1849

\bibitem[{Hodges-Kluck} \& {Bregman}(2013)]{HodgesBregman2013}
{Hodges-Kluck} E.~J., {Bregman} J.~N., 2013, \apj, 762, 12

\bibitem[{Hummels} et~al.(2013){Hummels}, {Bryan}, {Smith} \&
  {Turk}]{HummelsEtal2013}
{Hummels} C.~B., {Bryan} G.~L., {Smith} B.~D., {Turk} M.~J., 2013, \mnras, 430,
  1548

\bibitem[{Irwin} et~al.(2009){Irwin}, {Hoffman}, {Spekkens}
  et~al.]{IrwinEtal2009}
{Irwin} J.~A., {Hoffman} G.~L., {Spekkens} K., et~al., 2009, \apj, 692, 1447

\bibitem[{Joung} et~al.(2012{\natexlab{a}}){Joung}, {Bryan} \&
  {Putman}]{JoungEtal2012}
{Joung} M.~R., {Bryan} G.~L., {Putman} M.~E., 2012{\natexlab{a}}, \apj, 745,
  148

\bibitem[{Joung} et~al.(2012{\natexlab{b}}){Joung}, {Putman}, {Bryan},
  {Fern{\'a}ndez} \& {Peek}]{JoungEtal2012a}
{Joung} M.~R., {Putman} M.~E., {Bryan} G.~L., {Fern{\'a}ndez} X., {Peek}
  J.~E.~G., 2012{\natexlab{b}}, \apj, 759, 137

\bibitem[{Kalberla} \& {Dedes}(2008)]{2008A&A...487..951K}
{Kalberla} P.~M.~W., {Dedes} L., 2008, \aap, 487, 951

\bibitem[{Katz} et~al.(1996){Katz}, {Weinberg} \& {Hernquist}]{KWH1996}
{Katz} N., {Weinberg} D.~H., {Hernquist} L., 1996, \apjs, 105, 19

\bibitem[{Kaufmann} et~al.(2009){Kaufmann}, {Bullock}, {Maller}, {Fang} \&
  {Wadsley}]{KaufmannEtal2009}
{Kaufmann} T., {Bullock} J.~S., {Maller} A.~H., {Fang} T., {Wadsley} J., 2009,
  \mnras, 396, 191

\bibitem[{Kaufmann} et~al.(2006){Kaufmann}, {Mayer}, {Wadsley}, {Stadel} \&
  {Moore}]{KaufmannEtal2006}
{Kaufmann} T., {Mayer} L., {Wadsley} J., {Stadel} J., {Moore} B., 2006, \mnras,
  370, 1612

\bibitem[{Kaufmann} et~al.(2007){Kaufmann}, {Mayer}, {Wadsley}, {Stadel} \&
  {Moore}]{KaufmannEtal2007}
{Kaufmann} T., {Mayer} L., {Wadsley} J., {Stadel} J., {Moore} B., 2007, \mnras,
  375, 53

\bibitem[{Kere{\v s}} et~al.(2009){Kere{\v s}}, {Katz}, {Fardal}, {Dav{\'e}} \&
  {Weinberg}]{KeresEtal2009}
{Kere{\v s}} D., {Katz} N., {Fardal} M., {Dav{\'e}} R., {Weinberg} D.~H., 2009,
  \mnras, 395, 160

\bibitem[{Kere{\v s}} et~al.(2005){Kere{\v s}}, {Katz}, {Weinberg} \&
  {Dav{\'e}}]{KeresEtal2005}
{Kere{\v s}} D., {Katz} N., {Weinberg} D.~H., {Dav{\'e}} R., 2005, \mnras, 363,
  2

\bibitem[{Kim} et~al.(2014){Kim}, {Abel}, {Agertz} et~al.]{AGORApaper}
{Kim} J.-h., {Abel} T., {Agertz} O., et~al., 2014, \apjs, 210, 14

\bibitem[{King} et~al.(2008){King}, {Pringle} \& {Hofmann}]{KingEtal2008}
{King} A.~R., {Pringle} J.~E., {Hofmann} J.~A., 2008, \mnras, 385, 1621

\bibitem[{King} et~al.(2011){King}, {Zubovas} \& {Power}]{KingEtal2011}
{King} A.~R., {Zubovas} K., {Power} C., 2011, \mnras, 415, L6

\bibitem[{Lada} \& {Lada}(2003)]{LadaLada2003}
{Lada} C.~J., {Lada} E.~A., 2003, \araa, 41, 57

\bibitem[{Lee} \& {Chen}(2009)]{LeeChen2009}
{Lee} H.-T., {Chen} W.~P., 2009, \apj, 694, 1423

\bibitem[{Maller} \& {Bullock}(2004)]{MallerBullock2004}
{Maller} A.~H., {Bullock} J.~S., 2004, \mnras, 355, 694

\bibitem[{Maller} \& {Dekel}(2002)]{MallerDekel2002}
{Maller} A.~H., {Dekel} A., 2002, \mnras, 335, 487

\bibitem[{Marasco} et~al.(2012){Marasco}, {Fraternali} \&
  {Binney}]{MarascoEtal2012}
{Marasco} A., {Fraternali} F., {Binney} J.~J., 2012, \mnras, 419, 1107

\bibitem[{Marasco} et~al.(2013){Marasco}, {Marinacci} \&
  {Fraternali}]{MarascoEtal2013}
{Marasco} A., {Marinacci} F., {Fraternali} F., 2013, \mnras, 433, 1634

\bibitem[{Marinacci} et~al.(2011){Marinacci}, {Fraternali}, {Nipoti}, {Binney},
  {Ciotti} \& {Londrillo}]{MarinacciEtal2011}
{Marinacci} F., {Fraternali} F., {Nipoti} C., {Binney} J., {Ciotti} L.,
  {Londrillo} P., 2011, \mnras, 415, 1534

\bibitem[{Mashchenko} et~al.(2008){Mashchenko}, {Wadsley} \&
  {Couchman}]{MashchenkoEtal2008}
{Mashchenko} S., {Wadsley} J., {Couchman} H.~M.~P., 2008, Science, 319, 174

\bibitem[{Morganti} et~al.(2006){Morganti}, {de Zeeuw}, {Oosterloo}
  et~al.]{MorgantiEtal2006}
{Morganti} R., {de Zeeuw} P.~T., {Oosterloo} T.~A., et~al., 2006, \mnras, 371,
  157

\bibitem[{Nguyen Luong} et~al.(2011){Nguyen Luong}, {Motte}, {Hennemann}
  et~al.]{LuongEtal2011}
{Nguyen Luong} Q., {Motte} F., {Hennemann} M., et~al., 2011, \aap, 535, A76

\bibitem[{Ntormousi} et~al.(2011){Ntormousi}, {Burkert}, {Fierlinger} \&
  {Heitsch}]{NtormousiEtal2011}
{Ntormousi} E., {Burkert} A., {Fierlinger} K., {Heitsch} F., 2011, \apj, 731,
  13

\bibitem[{Pisano} et~al.(2007){Pisano}, {Barnes}, {Gibson}, {Staveley-Smith},
  {Freeman} \& {Kilborn}]{PisanoEtal2007}
{Pisano} D.~J., {Barnes} D.~G., {Gibson} B.~K., {Staveley-Smith} L., {Freeman}
  K.~C., {Kilborn} V.~A., 2007, \apj, 662, 959

\bibitem[{Pon} et~al.(2014){Pon}, {Johnstone}, {Bally} \&
  {Heiles}]{PonEtal2014}
{Pon} A., {Johnstone} D., {Bally} J., {Heiles} C., 2014, \mnras, 441, 1095

\bibitem[{Power} et~al.(2011){Power}, {Nayakshin} \& {King}]{PowerEtal2011}
{Power} C., {Nayakshin} S., {King} A., 2011, \mnras, 412, 269

\bibitem[{Prochaska} et~al.(2011){Prochaska}, {Weiner}, {Chen}, {Mulchaey} \&
  {Cooksey}]{ProchaskaEtal2011}
{Prochaska} J.~X., {Weiner} B., {Chen} H.-W., {Mulchaey} J., {Cooksey} K.,
  2011, \apj, 740, 91

\bibitem[{Read} \& {Hayfield}(2012)]{2012MNRAS.422.3037R}
{Read} J.~I., {Hayfield} T., 2012, \mnras, 422, 3037

\bibitem[{Read} et~al.(2010{\natexlab{a}}){Read}, {Hayfield} \&
  {Agertz}]{2009arXiv0906.0774R}
{Read} J.~I., {Hayfield} T., {Agertz} O., 2010{\natexlab{a}}, \mnras, 405, 1513

\bibitem[{Read} et~al.(2010{\natexlab{b}}){Read}, {Hayfield} \&
  {Agertz}]{ReadEtal2010}
{Read} J.~I., {Hayfield} T., {Agertz} O., 2010{\natexlab{b}}, \mnras, 405, 1513

\bibitem[{Rocha-Pinto} et~al.(2000){Rocha-Pinto}, {Scalo}, {Maciel} \&
  {Flynn}]{Rocha-PintoEtal2000}
{Rocha-Pinto} H.~J., {Scalo} J., {Maciel} W.~J., {Flynn} C., 2000, \apjl, 531,
  L115

\bibitem[{Saintonge} et~al.(2008){Saintonge}, {Giovanelli}, {Haynes}
  et~al.]{SaintongeEtal2008}
{Saintonge} A., {Giovanelli} R., {Haynes} M.~P., et~al., 2008, \aj, 135, 588

\bibitem[{Salpeter}(1955)]{Salpeter1955}
{Salpeter} E.~E., 1955, \apj, 121, 161

\bibitem[{Sanchez-Blazquez} et~al.(2014){Sanchez-Blazquez}, {Rosales-Ortega},
  {Mendez-Abreu} et~al.]{SanchezEtal2014}
{Sanchez-Blazquez} P., {Rosales-Ortega} F., {Mendez-Abreu} J., et~al., 2014,
  ArXiv e-prints

\bibitem[{Schmidt}(1959)]{Schmidt1959}
{Schmidt} M., 1959, \apj, 129, 243

\bibitem[{Schmidt}(1963)]{Schmidt1963}
{Schmidt} M., 1963, \apj, 137, 758

\bibitem[{Sommer-Larsen} et~al.(2003){Sommer-Larsen}, {G{\"o}tz} \&
  {Portinari}]{Sommer-LarsenEtal2003}
{Sommer-Larsen} J., {G{\"o}tz} M., {Portinari} L., 2003, \apj, 596, 47

\bibitem[{Springel}(2005)]{Springel05}
{Springel} V., 2005, \mnras, 364, 1105

\bibitem[{Stinson} et~al.(2006){Stinson}, {Seth}, {Katz}, {Wadsley},
  {Governato} \& {Quinn}]{2006MNRAS.373.1074S}
{Stinson} G., {Seth} A., {Katz} N., {Wadsley} J., {Governato} F., {Quinn} T.,
  2006, \mnras, 373, 1074

\bibitem[{Tumlinson} et~al.(2011){Tumlinson}, {Thom}, {Werk}
  et~al.]{TumlinsonEtal2011}
{Tumlinson} J., {Thom} C., {Werk} J.~K., et~al., 2011, Science, 334, 948

\bibitem[{Woolf} \& {West}(2012)]{WoolfWest2012}
{Woolf} V.~M., {West} A.~A., 2012, \mnras, 422, 1489

\bibitem[{Woosley} \& {Heger}(2007)]{WoosleyHeger2007}
{Woosley} S.~E., {Heger} A., 2007

\bibitem[{Worthey} et~al.(1996){Worthey}, {Dorman} \& {Jones}]{WortheyEtal1996}
{Worthey} G., {Dorman} B., {Jones} L.~A., 1996, \aj, 112, 948

\end{thebibliography}

\appendix

\section{Derivation of fit to low-J slope}

\begin{figure}
\begin{minipage}[b]{.49\textwidth}
\centerline{\psfig{file=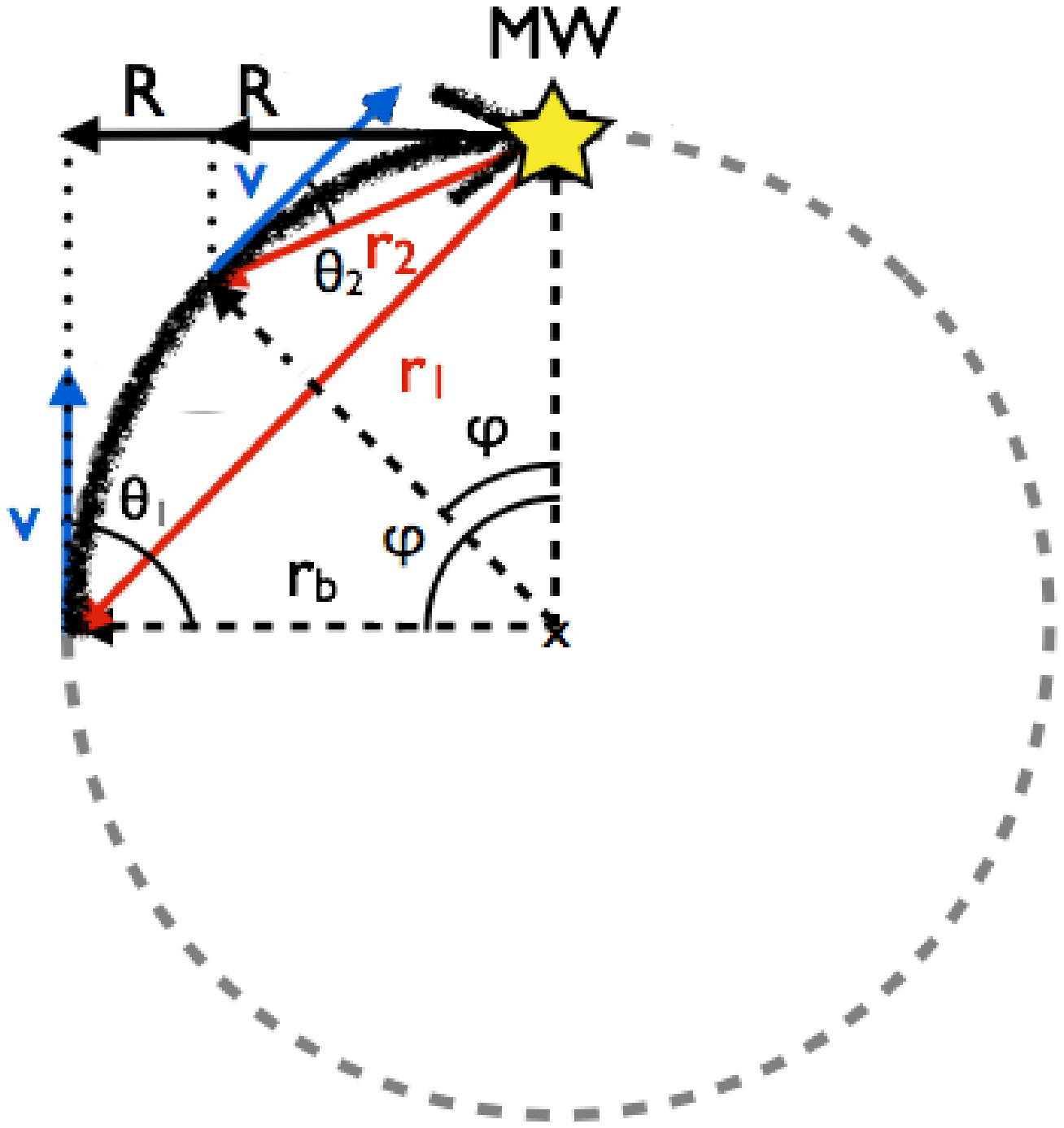,width=1.0\textwidth,angle=0}}
\end{minipage}
\caption[]{Schematic showing the condensation of a filament along the curved wall of a SNe bubble, used for the derivation of the low-J tail of the angular momentum distribution in Section \ref{sec:lowL}.}
\label{fig:cartoon}
\end{figure}

We consider two points on the circular arc that represents a `stream' condensing along the arc within the potential well. As per Figure \ref{fig:cartoon}, the circular arcs have radii of curvature $r_b$, the radius of the supernova bubble that creates the streams. The two locations on the arc have radii $r_1$ and $r_2$ from the centre of the potential, and projected radii $R_1$ and $R_2$ onto the axis defined along $x$ from where the bubble meets the origin. The tangential velocities at each point are $v_1$ and $v_2$. The angles created by the radii of curvature to each location are $\psi_1$ and $\psi_2$, and the angles between the velocity and radius vectors are $\theta_1$ and $\theta_2$. From geometry we can see that for each location we have:
\begin{equation}
\phi_1 = \sin^{-1} \left(\frac{R_1}{r_b}\right)
\end{equation}
and:
\begin{equation}
\phi_2 = \sin^{-1} \left(\frac{R_2}{r_b}\right)
\end{equation}
along with:
\begin{equation}
\theta_1 = \frac{\phi_1}{2}
\end{equation}
and:
\begin{equation}
\theta_2 = \frac{\phi_2}{2}
\end{equation}

The angular momenta per unit mass at each location are given by:
\begin{equation}
j_1 = r_1 v_1 \sin \theta_1
\end{equation}
and:
\begin{equation}
j_2 = r_2 v_2 \sin \theta_2
\end{equation}

The relations for each point along the circular arc (`stream') are identical, so in general for the entire stream we can write:
\begin{equation}
\theta = \sin^{-1} \left(\frac{r}{2r_b}\right)
\end{equation}
and so for the angular momentum per unit mass we have:
\begin{equation}
j = r v \sin \left[\sin^{-1} \left(\frac{r}{2r_b}\right)\right]
\end{equation}
giving:
\begin{equation}
j = r v \frac{r}{2r_b}
\end{equation}

Substituting in for the velocity as the free-fall velocity, $v_{\rm ff} = (2GM(r)/r)^{1/2}$, we have:
\begin{equation}
j = r \left(\frac{2 GM(r)}{r}\right)^{1/2} \frac{r}{2r_b}
\end{equation}
For $r \approx r_s$, the mass profile is largely isothermal, namely $M(r) \propto r$, although as $r$ becomes $\ll r_s$ this transitions to $M(r) \propto r^{20/9}$. In general, the streams span scales from $r \sim r_s$ down to $r \sim 5$ kpc where the galaxy is forming. Therefore, it seems reasonable to take an isothermal slope for $M(r)$ in this calculation. We note that for a stream that may extend down to smaller scales, e.g. inside the galaxy or even further down towards the SMBH, this slope would change.

So we have, using $M(r) = Ar$ where A is a proportionality constant,
\begin{equation}
j = r (2G)^{1/2} A^{1/2} \frac{r}{2r_b}
\end{equation}
and can therefore write, for the specific angular momentum of the cold filamentary gas, a proportionality of the form
\begin{equation}
j \propto r^2
\end{equation}
in contrast to the $j \propto r$ of the ambient gas set by the velocity field of the IC (refer to equation \ref{eq:ic}).

Substituting this back into our expression for $M$ in terms of $j$, we have for the low-angular momentum tail of the stream gas distribution
\begin{equation}
M \propto \left[\left(\frac{j^{1/2}}{r_s}\right)^{4/9}\right]^5
\end{equation}
and therefore
\begin{equation}
M \propto j^{1.1}
\end{equation}
i.e., a slope of approximately unity in the log-log histogram.

It is important to make the distinction between a stream that is \emph{condensing} along the paths discussed above and one where the gas is infalling from a single point of origin. In the latter case the primary conservation law is that of angular momentum, and so the value $j_{\rm initial}$ that the gas starts with will naturally be the value it keeps during its orbit, and the stream will have a single-valued $j$. For gas that is condensing along this path, however, we are not required to consider an orbit, and thus the geometry of the path can provide an angular momentum trend with radius. 

\end{document}